\documentclass[
prb,
reprint,
aps,
showpacs,
floatfix,
]{revtex4-1}

\usepackage{subcaption}
\usepackage{graphicx}
\usepackage[utf8]{inputenc}
\usepackage[T1]{fontenc}
\usepackage{amsmath}
\usepackage{amsfonts}
\usepackage{amssymb}
\usepackage{xcolor,colortbl}
\usepackage{dcolumn}
\usepackage{multirow}
\definecolor{grey}{RGB}{220,220,220}
\definecolor{matt}{RGB}{192,225,215}
\usepackage{todonotes}
\usepackage[percent]{overpic}
\usepackage{braket}
\usepackage{float}
\usepackage{cleveref}
\usepackage[justification=justified]{caption}

\makeatletter
\newcommand*{\addFileDependency}[1]{
  \typeout{(#1)}
  \@addtofilelist{#1}
  \IfFileExists{#1}{}{\typeout{No file #1.}}
}
\makeatother

\newcommand{\numcompounds}[0]{1260\ }

\newcommand{\mb}[1]{\mathbf{#1}}

\newcommand{\fro}{{Fr\"ohlich~}}
\def\olo	{{\omega_\mathrm{LO}}}
\def\ojlo	{{\omega_{j\mathrm{LO}}}}
\def\pjlo	{{p_{j\mathrm{LO}}}}
\def\meff	{{m^*}}
\def\H	{{\hat H}}
\def\kk	{{\mathbf{k}}}
\def\qq	{{\mathbf{q}}}
\def\pp	{{\mathbf{p}}}
\def\rr	{{\mathbf{r}}}
\def\ga  {{\alpha}}

\def\dd  {{\mathrm{d}}}
\def\KK {\,\mathrm{K}}

\def\aP {\,a_\mathrm{P}}
\def\zprs {$\mathrm{ZPRs}$}
\def\zprg {$\mathrm{ZPR_g}$}

\def\Cs {{Cs$_2$NaScF$_6$ }}

\begin{document}

\title{High-throughput analysis of Fr\"ohlich-type polaron models}
\author{Pedro Miguel M. C. de Melo$^{1,2}$}
\author{Joao C. de Abreu$^2$}
\author{Bogdan Guster$^3$}
\author{Matteo Giantomassi$^3$}
\author{Zeila Zanolli$^{1}$}
\author{Xavier Gonze$^{3}$}
\author{Matthieu J. Verstraete$^2$}

\affiliation{$^1$Chemistry Department, Debye Institute for Nanomaterials Science and European Theoretical Spectroscopy Facility, Condensed Matter and Interfaces, Utrecht University, PO Box 80.000, 3508 TA Utrecht, The Netherlands}
\affiliation{$^2$nanomat/Q-MAT/CESAM and European Theoretical Spectroscopy Facility, Universit\'e de Li\`ege, B-4000 Li\`ege, Belgium}
\affiliation{$^3$UCLouvain, Institute of Condensed Matter and Nanosciences (IMCN), Chemin des \'Etoiles~8, B-1348 Louvain-la-Neuve, Belgium}

\date{\today}

\begin{abstract}
The electronic structure of condensed matter can be significantly affected by the electron-phonon interaction, which leads to important phenomena such as electrical resistance, superconductivity or the formation of polarons.
This interaction is often neglected in band structure calculations, but can have a strong impact, e.g. on band gaps or optical spectra. 
Commonly used frameworks for electron-phonon energy corrections are the Allen-Heine-Cardona theory and the \fro model. 
The latter accounts for a single longitudinal optical mode, a single parabolic electron band, and washes out atomic details.
While it shows qualitative agreement with experiment for many polar materials, its simplicity should bring hard limits to its applicability in real materials.
Improvements can be made by introducing a generalized version of the model, which takes into account anisotropic and degenerate electronic bands, and multiple phonon branches. 
In this work, we search for trends and outliers on over a thousand materials in existing databases of phonon and electron band structures. 
We use our results to identify the limits of applicability 
of the standard \fro model by comparing to 
the generalized version, and by testing its basic hypothesis of 
a large radius for the polaronic wavefunction and the corresponding atomic displacement cloud (large polaron).
The validity of the perturbative approach to the \fro model is also tested. 
Among our extended set of materials, most exhibit large polaron behavior as well as validity of the perturbative treatment.
However, especially for the valence band, there is also a significant fraction of the materials for which the perturbative treatment cannot be applied and/or for which the size of the self-trapping region is close to the
atomic repetition distance.
We find a large variety of behaviors, and employ much more accurate, fully ab initio Allen-Heine-Cardona calculations to understand extreme cases, where the \fro model should fail and unusually large zero-point renormalization energies occur. 

\end{abstract}

\maketitle

\section{Introduction}
\label{intro}
The correct assessment of the electronic band gap and properties of charge carriers is of primary importance in determining the utility and applicability of semiconductors and insulators. 
Theoretical treatments usually only include the ``frozen-ion'' electronic aspect of the problem. 
Over the past two decades it has become clear that this is a severe limitation given the accuracy of both measurements and more advanced theory.\cite{Cardona2005,Shishkin2007,Marini2008,Giustino2010,Moser2013,Antonius2014,Ponce2015,Verdi2017,Verdi2017,Miglio2020}

The most common band-gap calculations involve Kohn-Sham Density Functional Theory (KS-DFT)\cite{HohKoh1964,KohSha1965,Martin2004} or the GW approximation from Many-body Perturbation Theory (MBPT), including different degrees of accuracy in the interactions between electrons\cite{Hedin1965,Martin2016}. 
MBPT computations are more computationally demanding than KS-DFT, but can yield band-gap results that are within  2\% to 10\% of experimental measurements\cite{Shishkin2007}. 
However, both are zero-temperature formalisms: a crucial and often ignored effect is the electron-phonon interaction (EPI), which leads to a renormalization of the band-gap as a function of temperature. 
Even at $T=0\KK$, EPI yields the so called zero-point renormalization of the band gap (ZPR$_{\rm c+v}$), that combines conduction and valence bands renormalizations (ZPR$_{\rm c}$ and ZPR$_{\rm v}$). 

Several theoretical approaches are available to calculate the ZPR$_{\rm c+v}$, among which the \fro model\cite{Frohlich1952} and the perturbative formalism proposed in the Allen-Heine-Cardona (AHC) approach\cite{Allen1976, Allen1981, Allen1983}.
In its first-principles version, AHC is the current gold standard for obtaining 
the ZPR$_{\rm c+v}$,\cite{Marini2008,Giustino2010,Antonius2014,Ponce2015,Verdi2015,Giustino2017,Sio2019,Gonze2020,Neaton2020} 
although its computational load is quite large.
In order for the AHC approach to be valid, the EPI should not be too strong, since it relies on a perturbative treatment.

In the original \fro model, the charge carrier dynamics is described by a one-band isotropic and parabolic dispersion, and couples to one dispersionless longitudinal optical phonon mode. 
The EPI is accounted for in a rather coarse fashion with a fixed analytic functional form, thanks to the hypothesis that the electron-phonon interaction is dominated by the long-range behavior of the Coulomb interaction, in effect washing out all atomic details. 
Studies of this model have been numerous\cite{Devreese2007,Feynman1955,Mishchenko2000,Vasilchenko2022}, and, depending on the EPI strength, can be done by perturbative means (weak coupling limit) or by a self-consistent approach to electron self-trapping 
by the phonon field (strong coupling limit).
Some techniques allow to cover the entire coupling strength range,\cite{Feynman1955,Mishchenko2000} but are either
difficult to generalize to first-principles approaches or require enormous computational resources.

The quasi-particle formed by a charge carrier dressed with phonons is called a ``polaron". 
Usually, the \fro model is only considered for so-called ``large'' polarons, for which the atomic details are ignored, while the denomination ``small polarons'' corresponds to the case where the localization of the electronic wavefunction is comparable to interatomic distances.
Interestingly, large polarons can be self-trapped as well, but 
in this case the self-trapping region is much larger than the interatomic distance. 

More recently Miglio et al. \cite{Miglio2020} derived a generalized \fro model, capturing a more realistic physical picture than the standard \fro model, in which one accounts for anisotropic and/or degenerate electronic dispersion, coupled to multiple phonons modes, possibly anisotropic, but still preserving the intrinsic continuum hypothesis (i.e. long-wavelength limit). 

While for the \fro model and its generalization only the zone-center phonons are needed, the AHC formalism requires the full phonon spectrum over the whole Brillouin Zone (BZ), and involves the explicit calculation of EPI matrix elements, making it computationally much more costly. 
ZPR$_{\rm c+v}$ determined via the generalized \fro model have shown comparable results to the AHC formalism for a set of materials that include oxides and II-VI compounds. However for less ionic materials, its predictions are not on a par with AHC.\cite{Miglio2020}

The materials studied in this work will be primarily polar, thus the use of the (generalized) \fro model is natural for two reasons: they are expected to yield polarons, and the model is a much less computationally costly estimation of the ZPR$_{\rm c+v}$. 
Given the recent developments on the \fro model\cite{Miglio2020} and the prevalence of polarons in different classes of semiconducting materials\cite{Franchini2021}, a thorough evaluation of the standard and generalized \fro~models over a broad range of materials is essential in order to establish each model's validity and limiting behaviors. 
One intrinsic aspect to the \fro model is the continuum limit, i.e. ignoring the crystal details while assuming that the polaron wavefunction is much larger than the shortest distance between atoms: the assumption is that one deals with large polarons. Other models are more suitable to include some level of atomistic detail, such as the Holstein model\cite{Holstein1959a,Holstein1959b}, which is not discussed here.
Independently of the length-scale aspect of the polaron problem, a qualitative criterion arises within the original \fro model with coupling strength $\alpha$.
In a weak-coupling perturbative treatment of this model, at $\alpha$ $\approx 6$ a breakdown occurs with the divergence of the effective mass.
This is to say that beyond this point simple perturbation theories fail, and the polaron experiences an intermediate or strong coupling with the crystal lattice deformations. 
Below this qualitative limit, in the weak coupling limit ($\alpha < 6$), the straight perturbative approach to the \fro model is in reasonably good agreement with more refined approaches such as Feynman's path integral variational approach\cite{Feynman1955} or diagrammatic Monte Carlo\cite{Mishchenko2000}. 
If the \fro model for a given material points to a breakdown of the perturbative approach, it is likely that its AHC treatment is also bound to fail, since it is based on a similar perturbative hypothesis.

The overall goal of this work is to exploit existing datasets from the literature to evaluate the breadth of applicability of the \fro model(s).
The development of high throughput workflows and database Application Programming Interfaces allows for fast queries of available information, allowing one to perform quick higher level calculations and even train machine learning algorithms. 
In this work, we rely on the database from Ref.~\onlinecite{Petretto2018}, which provides the electronic band structure, geometry, dielectric tensors, and phonon properties (e.g. mode frequencies, eigendisplacements) for a set of 1521 semiconducting materials. These were selected according to the following criteria: from two to five chemical elements per unit cell; experimentally stable 3D structures; non-magnetic; insulating materials with a minimal DFT band gap. The missing data on band masses is computed via a high throughput computational flow employing both AbiPy and ABINIT as described in Sec.~\ref{sec:eff_mass}. In the end, \numcompounds materials had all needed quantities to parameterize a \fro model Hamiltonian. The remaining 261 materials either have unstable phonon modes or their band extrema were not located along high-symmetry lines in the BZ. The latter issue makes automating the computation of band effective masses extremely difficult, as one has to find the global maximum (minimum) of the valence (conduction) band in the BZ.
We do not believe that their inclusion in this work would significantly alter the results shown herein. 

We focus on indicators that could mark the potential of a material as a system with large or small polarons. These may be desirable (for optical properties) or undesirable (for transport) in different applications: more is not necessarily better.
The essential quantity in the standard \fro model $\ga$, and the parameters of its generalized form~\cite{Miglio2020},  involve the dielectric tensor, the effective masses at the band extrema, the Born effective charge tensor, and the phonon frequencies, all of which are stored in the databases mentioned above. 

The paper is structured as follows: in Sec. \ref{sec:theory}A we recap the theoretical background for the original \fro model in describing large polarons, while in Sec. \ref{sec:theory}B we summarize the recent developments of the generalized \fro model. 
We present the high-throughput results for the \numcompounds studied materials in Sec. \ref{sec:highthroughput}. 
We follow with Sec. \ref{sec:abichecks} by performing an $ab~initio$ validation using the AHC approach for selected materials, and we discuss outliers found using the \fro model, such as materials with large ZPR and small coupling $\alpha$, as well as materials with large ZPR and large $\alpha$. In these cases the \fro model should be treated in the strong-coupling limit, and the long-wavelength limit may fail entirely (these are two distinct cases).
We conclude in Sec. \ref{sec:conclusions}.

\section{Theory}
\label{sec:theory}

\subsection{The standard \fro model}
\label{fro-model}
The \fro model~\cite{Frohlich1952,Frohlich1954} assumes a system with a single parabolic electron band of effective isotropic mass $\meff$ and a single non-dispersive longitudinal optical phonon branch of frequency $\olo$. 
The electron-phonon interaction comes from the macroscopically screened Coulomb interaction between electrons and the nuclei moving along the optical phonon mode. 
While the latter approximation is dominant and qualitatively correct for $\qq\approx0$, it is assumed to be valid in the whole Brillouin zone, which
corresponds to a continuum treatment, in line with the hypotheses of an isotropic electronic band and non-dispersive phonons. 
This means that there are no Debye-Waller contributions to the electron-phonon interaction, and transverse optical or acoustic modes are ignored. 
This model also ignores band degeneracies and the possibility of different band masses and warping.\cite{Mecholsky2014}
Although formulated initially for a conduction electron, it can be easily applied to valence electrons, with a proper change of sign in selected formulas.

For a material with non-degenerate isotropic band extrema and isotropic dielectric function, we can write the following Hamiltonian for the standard \fro model for an electron (conduction band - in atomic units $\hbar$=1, $a_{\rm Bohr}$=1, $m$=1).\cite{Mahan2000}

\begin{multline}
	\H^\mathrm{sFr} =  \sum_\kk \frac{\kk^2}{2\meff}\hat c_\kk^\dagger \hat c_\kk + \sum_\qq\olo\hat a_\qq^\dagger\hat a_\qq +\\
	 \sum_{\kk.\qq} g^\mathrm{sFr}(\qq)\hat c_{\kk+\qq}^\dagger \hat c_\kk \left(\hat a_\qq^\dagger + \hat a_{-\qq} \right),
\label{eq-hfro}
\end{multline}
where the electron-phonon coupling constant is given by
\begin{equation}
	g^\mathrm{sFr}(\qq) = \frac{1}{q} \left(\frac{2\pi\olo}{\epsilon^*V_\mathrm{BvK}} \right)^\frac{1}{2}
	= \frac{1}{q} \left(\frac{2\sqrt{2}\pi}{V_\mathrm{BvK}}\frac{\olo^{3/2}}{\sqrt{\meff}}\ga\right)^\frac{1}{2}, 
\end{equation}
$V_\mathrm{BvK}$ is the Born von K\'arm\'an supercell volume, the dimensionless coupling parameter $\ga$ is 
\begin{equation}
    \ga = \frac{1}{\epsilon^*} \sqrt{\frac{\meff}{2\olo}},
    \label{eq-alpha}
\end{equation}
where $\epsilon^*$ is defined by 
\begin{equation}
\frac{1}{\epsilon^*}=
\frac{1}{\epsilon^\infty}
-
\frac{1}{\epsilon^0}.
\label{eq-epsilon-1-star}
\end{equation}
As a side note, $\epsilon^*$, $\epsilon^\infty$ and $\epsilon^0$ are independent of the nuclear masses: $\epsilon^\infty$ is purely electronic, while $\epsilon^0$ is obtained in the adiabatic limit (low-frequency limit), so that the nuclei have time to adjust adiabatically to the applied electric field, regardless of their mass. 
$\epsilon^*$ is always greater 
than $\epsilon^\infty$.
For polar compounds with strong ionic screening, the difference in $\epsilon$'s will be large, and $\epsilon^* \rightarrow \epsilon^\infty$ whereas for purely covalent compounds or mono-atomic compounds $\epsilon^0 \simeq \epsilon^\infty$, and so $\epsilon^* \rightarrow \infty$, which runs counter to intuition for habitual dielectric responses, but simply pushes $\ga$ to 0 in the \fro model.
In Eqs.~\eqref{eq-hfro} and~\eqref{eq-epsilon-1-star}, we follow the Born and Huang convention for the phonon eigenvectors at $\qq$ and $-\qq$.\cite{Guster2022}

The Hamiltonian in Eq.~\eqref{eq-hfro} can be simplified by using a different choice of units: 
\begin{multline}
	\H^\mathrm{sFr} =  \sum_\kk \frac{\kk^2}{2}\hat c_\kk^\dagger \hat c_\kk + \sum_\qq\hat a_\qq^\dagger\hat a_\qq +\\
	 \sum_{\kk.\qq} \frac{1}{q} \left(\frac{2\sqrt{2}\pi\ga}{V_\mathrm{BvK}} \right)^\frac{1}{2}\hat c_{\kk+\qq}^\dagger \hat c_\kk \left(\hat a_\qq^\dagger + \hat a_{-\qq} \right),
\label{eq-hfro-res}
\end{multline}
where the energies, momenta, and length were rescaled by factors of $\olo$, $(\olo\meff)^{1/2}$, and $(\olo\meff)^{-1/2}$, respectively. 
With this choice, it becomes clear that the sole free parameter $\ga$ characterizes the strength of the electron-phonon interaction with respect to the intrinsic electron and phonon terms.

Such a Hamiltonian can be treated by perturbation theory,
in the limit of small $\ga$, delivering the polaron binding energy (ZPR) 
(again in atomic units instead of $\olo$ units) as
\begin{equation}
	E_\mathrm{P}=-\olo(\ga+0.0159\ga^2+...).
	\label{eq-zpr}
\end{equation}
A more accurate approach based on the Feynman path integral\cite{Feynman1955}
can be employed, covering the whole range of coupling strengths, as follows:

\begin{equation}
    \begin{aligned}
    E_\mathrm{P} = -\olo( & \ga+0.98(\ga/10)^2+0.60(\ga/10)^3 \\
                          & +0.14(\ga/10)^4), \ga \leq 5 \\
    E_\mathrm{P} = -\olo( & 0.106\ga^2+2.83), \ga \geq 5.
    \end{aligned}
    \label{eq:zpr_fey}
\end{equation}

With the same perturbative treatment, it is possible to show that the ratio between the effective masses of the polaron, $m^*_\mathrm{P}$, and the electron is approximately given by
\begin{equation}
	\frac{m^*_\mathrm{P}}{\meff} = \left(1 - \frac{\ga}{6} + 0.00417\ga^2+...\right)^{-1}.
	\label{eq-polar-mass}
\end{equation}

At the lowest order of perturbation theory,
one obtains the well-known formula
\begin{equation}
	E_\mathrm{P}
	\approx
	-\ga\olo,
	\label{eq-zpr-pt}
\end{equation}
or more explicitly,
\begin{equation}
    E_\mathrm{P} = -\frac{1}{\epsilon^*} \sqrt{\frac{\meff\olo}{2}},
    \label{eq-EP}
\end{equation}
where $\olo$ contains the only dependence on nuclear masses.
At this order in the expansion Eq.~\eqref{eq-polar-mass} yields the following polaron mass:
\begin{equation}
	\frac{m^*_\mathrm{P}}{\meff} \approx \left(1 - \frac{\ga}{6}\right)^{-1}.
\end{equation}
with the immediate consequence that at $\ga=6$ the polaron mass diverges and this low-order perturbation theory approach is no longer valid. 
This parameter thus also provides a breakdown point for the lowest-order perturbative treatment of the \fro model. 
A large $\ga$ is physically associated with the appearance of self-localization of the electron due to the phonon response, a non-perturbative phenomenon that can be treated, alternatively, in the so-called ``strong-coupling'' limit of the \fro model.
Thus the $\ga=6$ value suggests a change of regime for the \fro polaron. Nevertheless the occurrence of the wide range of behaviors present in our set of \numcompounds materials
demands a more careful treatment in describing the polaron effective mass: we make use of the results based on the Diagrammatic Monte Carlo method applied to the standard \fro model\citep{Mishchenko2000}, by mapping the corresponding electronic and polaronic effective masses in the available range of $\ga$ (see Fig. 5 in Ref. \emph{\citenum{Mishchenko2000}}). Outside of the available range we fit a smooth and continuous quartic function with a resulting best fit as follows:

\begin{equation}
    \frac{m^*_\mathrm{P}}{\meff} = \left( 1.07\ga^4 - 160.53  \right)^{-1}.
    \label{eq:mp_dmc}
\end{equation}

The self-localisation of the electron yields the notion of a ``polaron radius'', $\aP$. For instance, with a Gaussian ansatz for the electronic wavefunction, in the adiabatic limit one obtains:
\begin{equation}
	\phi(\rr)= 
	\left( \frac{1}{\aP \sqrt{\pi}}
	\right)^{\frac{3}{2}}
	\exp 
	\left( -\frac{\rr^2}{2\aP^2}
	\right),
\end{equation}
with 
\begin{equation}
	\aP= 3\sqrt{\frac{\pi}{2}}
	\frac{\epsilon^*}{m^*}.
	\label{eq-ap-radius}
\end{equation}
Coherently, $\aP$ is defined only in terms of quantities that do not depend on the nuclear masses. 
When $\aP$ is on the order of the distance between equivalent atomic sites in the crystals, the \fro model cannot be a good representation of the real material, as it is based on a continuum hypothesis for the vibrational degrees of freedom. 

\subsection{The generalized \fro model}
\label{sec:gfro}
The \fro model can be generalized to include systems with degenerate and anisotropic band extrema, multiple phonon branches, and anisotropic dielectric functions~\cite{Miglio2020}. 
Bands are still assumed to be parabolic in each direction and phonon energies are still constant with respect to the wave vector length $\qq$, but all might depend on the direction.
As in Ref.~\onlinecite{Guster2021}, we treat both conduction and valence bands thanks to the integer variable $\sigma$, that is $1$ for the conduction band (or electron polarons), and $-1$ 
for the valence band (or hole polarons).
The Hamiltonian is then similar to that of Eq.~\eqref{eq-hfro},

\begin{multline}
	\H^\mathrm{gFr} =  \sum_{\kk n} \frac{\sigma\kk^2}{2m_{n}^{*}(\hat\kk)}\hat c_{\kk n}^\dagger \hat c_{\kk n} + \sum_{\qq j} \omega_{j0}(\hat\qq)\hat a_{\qq j}^\dagger\hat a_{\qq j} + \\\sum_{\qq j, \kk n'n} g^\mathrm{gFr}(\qq j,\kk n'n)\hat c_{\kk+\qq n'}^\dagger \hat c_{\kk n}\left(\hat a_{\qq j}^\dagger + \hat a_{-\qq j}\right),
\label{eq-ghfro}
\end{multline}
with $m^*_n(\hat{\mathbf{k}})$ the direction-dependent effective masses, $\mathbf{k}$ the electron wavevector,
 $n$ the band index, $\omega_{j0}(\hat \qq)$  the direction-dependent phonon frequency, $\qq$ the phonon wavevector and $j$ the phonon branch index.
The electron-phonon coupling constant is given by
\begin{eqnarray}
g^{\rm gFr}(\mathbf{q}j,\mathbf{k}n'n)=
\frac{i}{q} &&\frac{4\pi}{\Omega_0} 
\Biggl (
\frac{1}{2\omega_{j0}(\hat{\mathbf{q}})V_{\rm BvK}}
\Biggr )^{1/2}
\frac{\hat{\mathbf{q}} \cdot \mathbf{p}_j(\hat{\mathbf{q}})}
        {\epsilon^\infty(\hat{\mathbf{q}})}
\nonumber
\\
\times &&
\sum_m
s_{n'm}(\hat{\mathbf{k}}')
(s_{nm}(\hat{\mathbf{k}}))^*.
\label{eq:SM_ggFr}
\end{eqnarray}
In these equations, the sum over $n$, $n'$ and $m$ runs only over the bands that connect to the degenerate extremum, that are renumbered from 1 to $n_{\rm deg}$. 
The electron-phonon part also depends only on few quantities: the Born effective charges (entering the mode-polarity vectors $\mathbf{p}_j$ which are the Born charge weighted phonon displacement vectors), the macroscopic dielectric tensor $\epsilon^\infty$, and the phonon frequencies $\omega_{j0}$, the primitive cell volume $\Omega_0$, the Born-von Karman normalization volume $V_{\rm BvK}$ corresponding to the $\mathbf{k}$ and $\mathbf{q}$ samplings. The $s_{nm}$ tensors are symmetry-dependent unitary matrices, similar to spherical harmonics. Finally, $\kk'=\kk+\qq $.

In  this generalized model, the ZPR for a band extremum can be obtained also at lowest order of perturbation theory, as
\begin{multline}
	\mathrm{ZPR^{g F r}}=-\sum_{j n} \frac{\sigma}{\sqrt{2} \Omega_{0} n_{\rm deg}} \int_{4 \pi} \dd \hat{\mathbf{q}}\,\left(m_{n}^{*}(\hat{\mathbf{q}})\right)^{1 / 2}\times \\
	\left(\omega_{j 0}(\hat{\mathbf{q}})\right)^{-3 / 2}\left(\frac{\hat{\mathbf{q}} \cdot \mathbf{p}_{j}(\hat{\mathbf{q}})}{\epsilon^{\infty}(\hat{\mathbf{q}})}\right)^{2}.
	\label{eq-gzpr}
\end{multline}
When comparing with the expression for the renormalization energy from the standard \fro model, Eq.~\eqref{eq-zpr-pt}, we see that it is possible to re-write Eq.~\eqref{eq-gzpr} in a similar way, highlighting the fact that this expression originates from an average over $\hat{\qq}$ directions, and summation over the contributions from different phonon branches;
\begin{equation}
	\mathrm{ZPR}^\mathrm{gFr}= 
	- \sigma\sum_{j}
	\big \langle \, \ga_{j}(\hat{\qq}) 
	\omega_{j 0}(\hat{\qq})
	\big \rangle_{\hat{\qq}},
	\label{eq-gzpr2}
\end{equation}
where
\begin{eqnarray}
\big \langle \, f(\hat{\qq})
\big \rangle_{\hat{\qq}}
=\frac{1}{4 \pi} \int_{4 \pi} \dd \hat{\qq} \, f(\hat{\qq})
	\label{eq-avgq}
\end{eqnarray}
is an average over $\hat{\qq}$ directions.

The $\ga_j(\hat \qq)$ parameters are defined by
\begin{multline}
	\ga_{j}(\hat{\qq})=\frac{4 \pi}{\sqrt{2} \Omega_{0}}\left(\frac{1}{n_{\text {deg }}} \sum_{n=1}^{n_{\text {deg }}}\left(m_{n}^{*}(\hat{\qq})\right)^{1 / 2}\right)\times\\
\left(\omega_{j 0}(\hat{\qq})\right)^{-1 / 2}\left(\frac{\hat{\qq} \cdot \pp_{j}(\hat{\qq})}{\epsilon^{\infty}(\hat{\qq})\omega_{j 0}(\hat{\qq})}\right)^{2},
	\label{eq-ga1}
\end{multline}
and can also be re-written to look similar to Eq.~\eqref{eq-alpha},
\begin{equation}
	\ga_{j}(\hat{\qq})=
	\frac{
	\big \langle (m^*_n(\hat{\qq}))^{\frac{1}{2}}
	\big \rangle_n}
	{\epsilon_j^*(\hat{\qq})\sqrt{2\omega_{j 0}(\hat{\qq})}}
	\label{eq-ga1-jq}
\end{equation}
where
\begin{equation}
\frac{1}{\epsilon_j^*(\hat{\qq})}
=
\frac{4 \pi}{\Omega_{0}}
\left(\frac{\hat{\qq} \cdot \pp_{j}(\hat{\qq})}{\epsilon^{\infty}(\hat{\qq})\omega_{j 0}(\hat{\qq})}\right)^{2}
\label{eq-epsilon-1-star-q}
\end{equation}
replaces Eq.~\eqref{eq-epsilon-1-star}, while
\begin{equation}
\big \langle (m^*_n(\hat{\qq}))^{\frac{1}{2}}
	\big \rangle_n
	=
	\frac{1}{n_{\text {deg }}}
	\sum_{n=1}^{n_{\text {deg }}}\left(m_{n}^{*}(\hat{\qq})\right)^{1 / 2}
    \label{eq-m-start-q}
\end{equation}
highlights that the effective mass entering Eq.~\eqref{eq-ga1-jq} is an average over bands that are degenerate at the extremum.

To summarize, in the lowest order of perturbation theory treatment, the multiband, multibranch, anisotropic generalization of the simple Eq.~\eqref{eq-alpha} can be structured in the same way, with the band contribution being averaged, the branch contributions being summed, and the anisotropy being treated by an average over $\hat{\qq}$ directions. 
The polaron formation energy writes
\begin{equation}
    E_\mathrm{P} = -\sum_{j}
    \Big \langle
    \frac{\big \langle (m^*_n(\hat{\qq}))^{\frac{1}{2}}
	\big \rangle_n}{\epsilon_j^*(\hat{\qq})} \sqrt{\frac{\omega_{j 0}(\hat{\qq})}{2}}
    \Big \rangle_{\hat{\qq}}.
    \label{eq-EP-gen}
\end{equation}
Note that $\sigma$ does not appear in $E_\mathrm{P}$.

A generalization of $\alpha$ to anisotropic, multibranch systems, $\ga_{j}(\hat{\qq})$, has been tentatively defined by C. Verdi, see Ref.~\onlinecite{Verdi2017a}, Eq.(4.12) page 62, however lacking both effective mass and phonon frequency dependencies on direction, and ignoring the possible electronic degeneracy.

From Eq.~\eqref{eq-EP-gen} one might examine the relevance of the following approximate decoupling between electronic and vibrational and dielectric contributions :
\begin{eqnarray}
    E_\mathrm{P} \approx
    -\big \langle (m^*_n(\hat{\qq}))^{\frac{1}{2}}
	\big \rangle_{\hat{\qq}n}
	\Bigg( \sum_{j}
    \Big \langle
    \frac{1}{\epsilon_j^*(\hat{\qq})} \sqrt{\frac{\omega_{j 0}(\hat{\qq})}{2}}
    \Big \rangle_{\hat{\qq}} \Bigg).
\nonumber\\
    \label{eq-EP-gen-approx}
\end{eqnarray}
The factorization of the $\big \langle (m^*_n(\hat{\qq}))^{\frac{1}{2}}
	\big \rangle_{\hat{\qq}n}$ term appears naturally and is exact in the cubic case.

\subsection{The generalized \fro model in cubic systems}
\label{sec:gcubicfro}

For a cubic system, even with several phonon branches, there is no dependence on the direction of $\qq$ for the phonon frequencies and dielectric properties. The Hamiltonian becomes~\cite{Guster2021, Guster2022}
\begin{multline}
	\H^\mathrm{cFr} =  \sum_{\kk n} \frac{\sigma\kk^2}{2m_{n}^{*}(\kk)}\hat c_{\kk n}^\dagger \hat c_{\kk n} + \sum_{\qq j}\ojlo \hat a_{\qq j}^\dagger\hat a_{\qq j} + \\\sum_{\kk n'n,\qq j} g^\mathrm{cFr}(\qq j,\kk n'n)\hat c_{\kk+\qq n'}^\dagger \hat c_{\kk n}\left(\hat a_{\qq j}^\dagger + \hat a_{-\qq j}\right),
\label{eq-chfro}
\end{multline}
with a slightly simplified electron-phonon coupling constant given by
\begin{eqnarray}
g^{\rm cFr}(\qq j,\mathbf{k}n'n)=
\frac{1}{q} &&\frac{4\pi}{\Omega_0} 
\left(  \frac{1}{2\ojlo V_{\rm BvK}}
\right)^{1/2}
\frac{\pjlo}{\epsilon^\infty}
\nonumber
\\
\times &&
\sum_m
s_{n'm}(\hat{\mathbf{k}}')
(s_{nm}(\hat{\mathbf{k}}))^*.
\label{eq:SM_gcFr}
\end{eqnarray}
The suppression of the ``$i$'' prefactor from Eq.~\ref{eq:SM_ggFr} to Eq.~\eqref{eq:SM_gcFr} is related to the Born and Huang convention~\cite{Guster2022}.
Still working in the lowest order of perturbation theory, Eq.~\eqref{eq-gzpr2} can be re-written as a linear combination of $\ga_j$ parameters and phonon frequencies at $\Gamma$,
\begin{equation}
	\mathrm{ZPR^{cFr}}=-\sigma\sum_{j} \ga_{j} \ojlo,
	\label{eq-gzpr-cubic}
\end{equation}
with each $\ga_j$ being
\begin{equation}
	\ga_{j}= 
	\frac{\big \langle (m^*_n(\hat{\qq}))^{\frac{1}{2}}
	\big \rangle_{\hat{\qq}n}
	}
	{
	\epsilon_j^*\sqrt{2\ojlo}
	}
	\label{eq-ga1-j}
\end{equation}
]
The numerator is purely electronic, and independent of the $j$ index, while the denominator is purely dielectric and dynamical. This simplification appears only for the cubic crystallographic system. A similar decoupling appears for the polaron energy:
\begin{equation}
E_\mathrm{P}=
	-\big \langle (m^*_n(\hat{\qq}))^{\frac{1}{2}}
	\big \rangle_{\hat{\qq}n}
	\left( \sum_{j}
	\frac{\sqrt{\ojlo}}{\epsilon_j^*\sqrt{2}}
	\right).
	\label{eq-gzpr-cubic-simple}
\end{equation}

These equations shed light on the relationship between the standard and generalized \fro model.
They will provide guidance for the choice of the
parameters for the standard \fro model, which follows in the next section.

\subsection{Effective masses}
\label{sec:eff_mass}
An automatic python workflow was created to obtain the electronic effective masses and store them in a database. The input files for ABINIT~\cite{Gonze2020,Romero2020} were generated using AbiPy~\cite{Gonze2020,Romero2020} by inserting the ground-state parameters from the Materials Project database. 
The valence band maximum (VBM) and conduction band minimum (CBM) were then determined by AbiPy by producing the electronic band structure along the high-symmetry paths of each system's Brillouin zone. 
The effective mass tensors at these points were determined by calculating the second order derivative of the VBM and CBM eigenenergies with respect to the wavevector $\mb{k}$, including the effect of band degeneracies and warping \citep{Mecholsky2014}, within Density Functional Perturbation Theory (DFPT)\citep{Janssen2016}, as implemented in ABINIT. 

In section \ref{sec:highthroughput}, we use the directional average of effective masses,
\begin{equation}
    m^* \approx \left( \langle (m^*_n(\hat{\qq}))^{\frac{1}{2}} 
	\big \rangle_{\hat{\qq}n}
	\right)^2,
    \label{eq-sfr-mass}
\end{equation}
to facilitate the characterization of the materials. For cubic systems, this expression accounts exactly for the anisotropy in the effective masses
and for the possible degeneracies, as discussed in Sec.~\ref{sec:gcubicfro}.
For non-cubic systems, there is no such decoupling of the effective mass factor from the dielectric and dynamical ones, as 
Eq.\eqref{eq-EP-gen-approx} is not exact.

\section{High throughput results}
\label{sec:highthroughput}

\subsection{Standard \fro model}\label{sec:HT_standardmodel}
The standard \fro model relies on the $\olo$ frequency, the
$m^*$ effective mass and the ionic part of the dielectric tensor, 
$\epsilon^*$. 
They combine to deliver the $\ga$ parameter, and to predict a polaron formation energy $E_{\rm P}$ (or ZPR) obtained in the full range of
values from diagrammatic Monte Carlo or Feynman path integral approaches\cite{Feynman1955,Mishchenko2000}. 
If we want to define a standard \fro model for a given
material, we must extract a single LO phonon frequency, an isotropic dielectric tensor and a single mass, while this situation applies exactly only to the simplest materials, namely binary (and some ternary cubic) materials
with the electron or hole pockets situated at $\Gamma$.
The standard \fro model has nevertheless been widely used. 

In order to extract such simplified parameters in a high-throughput approach, and compare to the
generalized \fro model and the AHC first-principles treatment, we work with the following hypotheses.

For the dielectric constants we use $\epsilon = \mathrm{Tr}(\epsilon_{\ga\beta})/3$.
The single phonon frequency $\olo$ is assumed to be the highest phonon frequency
at $\Gamma$. 
For all the simple materials mentioned above, this is indeed the case (the highest frequency mode is always an LO). 
When several LO phonon modes are present, one could expect that the highest one has the largest LO-TO splitting and, thus, interacts most strongly with the
electrons.
We will see, however, that there are cases for which the highest mode is weakly coupled to electrons, and other LO modes are dominant (see the metal-azides below as an example of such a situation). 

\begin{figure}[!htb]
    \centering
    \includegraphics[width =0.9\columnwidth]{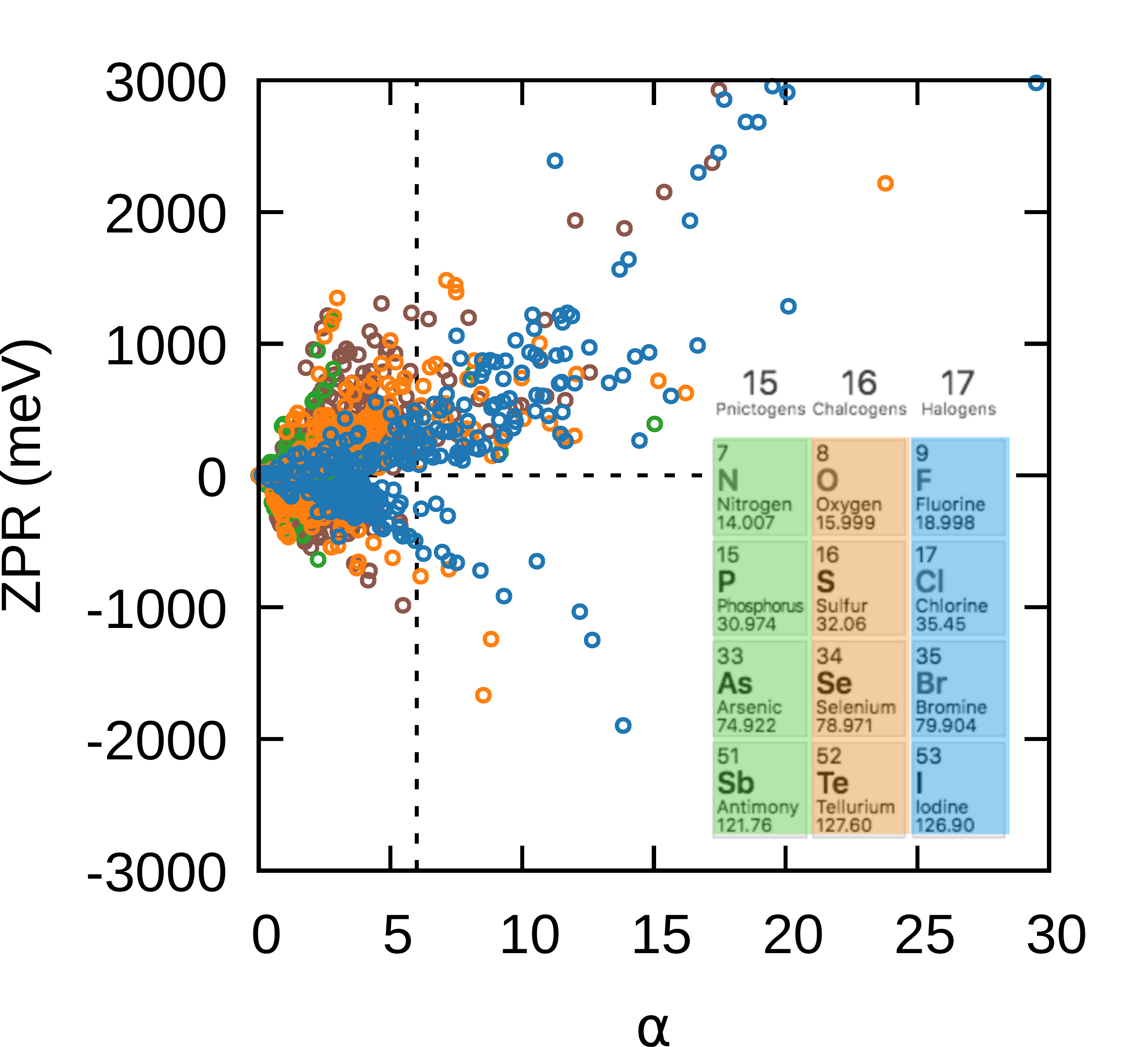}
    \caption{Conduction (negative) and valence (positive) standard \fro model ZPR and $\ga$ values, for all materials except 39 (1) exceeding 3000 meV for valence (conduction) ZPR. The ZPR values are determined based on the full range coupling strength described in Eq. \eqref{eq:zpr_fey}. The color corresponds to chemical elements from groups 15 to 17 of the periodic table (see inset), and brown for all other compounds. Most compounds that exhibit large values of $\ga$ have at least one element from group 17.
    Valence values are distributed over a wider range of $\ga$, and conduction values are more concentrated below $\ga = 10$. The vertical dashed  line is at $\ga = 6$. }
    \label{fig:szpr-2}
\end{figure}
As stated in Section~\ref{intro}, there are \numcompounds materials for which all necessary quantities are present to compute the ZPR$^\mathrm{sFr}$ and $\ga$ using Eqs.~\eqref{eq-alpha} and~\eqref{eq-zpr-pt}.
In Fig.~\ref{fig:szpr-2} we show the dispersion map of these quantities for both valence and conduction band edges. The color of each point indicates the presence of an element of a given group of the periodic table (as shown in the inset), according to the following order of precedence: blue for materials with an element from group 17 (halides); if no halide is present, orange for materials with elements from group 16 (chalcogenides); green for materials with an element from group 15 (pnictogens); red for materials with an element from group 14. If no element of any of these groups is present, the circle is brown. 

\begin{figure}[!htb]
    \centering
    \includegraphics[width =0.9\columnwidth]{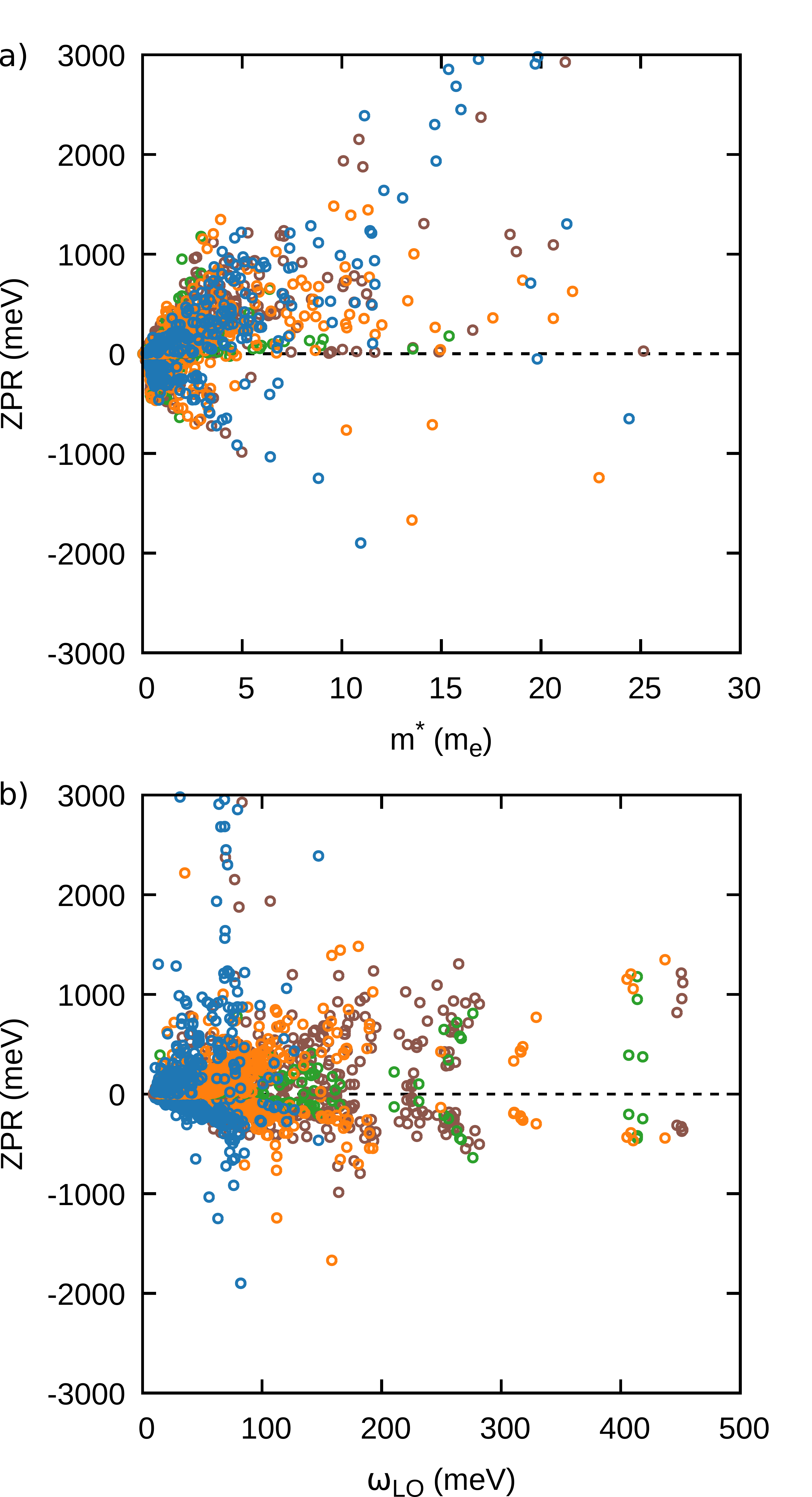}
\caption{Dispersion of conduction (negative) and valence (positive) standard ZPR energies versus the effective mass, $m^*$ (top), and phonon frequency, $\olo$ (bottom) for all materials with ZPR below 3000 meV. Same conventions as in Fig.~\ref{fig:szpr-2}. A rough square root behavior (Eq.~\eqref{eq-EP}) governs the maximum accessible ZPR for a given mass, and a degree of clustering is visible of the frequencies as a function of chemical period, with the lowest frequencies for halides, followed by chalcogens, then the remaining materials. Dependence with the band effective mass (Eq.~\eqref{eq-sfr-mass}) shows a dominant linear behaviour, with a wider dispersion for valence bands when compared to the conduction band masses.}
\label{fig:szpr-3}
\end{figure}

Values of $\ga$ for conduction states are almost entirely concentrated in the $\ga<10$ region, while valence values extend further into the $10<\ga<20$ range. 
This difference comes from the different distributions of bare effective masses for conduction and valence band edges\cite{Hautier2014} (shown in the top panel of Fig.~\ref{fig:szpr-3}), since $\epsilon^*$ and $\olo$ are the same for a given material. 
The ZPR distributions for both valence and conduction states follow a similar distribution, with few absolute values of conduction ZPR above 1000 meV, while the [1000,3000] meV range for absolute ZPR values is more populated in the valence case. 

The broad trend ZPR that is proportional to $\ga$ in a family is visible for both conduction and valence bands, coming from the simple proportionality through the $\olo$ frequency
in the lowest order of perturbation theory.
Not all chemical families show the same slope, even if the
group of constituting element is taken into account, because the nuclear mass and bonding vary within a group, which can strongly influence $\olo$ and $\epsilon^0$.

Halides (group 17) produce the highest values of $\ga$, far beyond the limit of validity of the perturbation treatment. 
Then the chalcogenides show the next set of large values of $\ga$, followed by the pnictogens and compounds with elements of group $\leq$ 14, with the lowest $\ga$. This trend is chemically intuitive as halides have a stronger polar behavior than elements of previous groups: polaron binding grows with electron affinity, but compounds from the ``other'' category, with no strongly electrophilic element, can be found in the full range up to quite high $\ga \sim 20$.

By looking at Eqs.~\eqref{eq-alpha},~\eqref{eq-epsilon-1-star}, and~\eqref{eq-zpr-pt} we split the descriptors into two categories: electronic properties with $1/\epsilon^\infty$ and $m^*$; and vibrational properties, $1/\epsilon^*$ and $\olo$. Of these, only $m^*$ and $\olo$ show clear clustering or trends, see Fig.~\ref{fig:szpr-3}.
The data shown in these figures allows to further understand the dispersion of values shown in Fig.~\ref{fig:szpr-2}. 
On the one hand, the highest values of $m^*$ and $1/\epsilon^\infty$ are obtained in halides, followed by chalcogens, which contributes to their large values of both $\ga$ and ZPR. 
The largest effective masses come from transition metal halides where the conduction band is an empty d-band, isolated due to crystal field splitting, which can become extremely flat (e.g. CaTiF$_6$).
Regarding the dependency of ZPR on $\omega_{\rm LO}$ shown in Fig.~\ref{fig:szpr-3} b), two points must be mentioned. 
The first concerns the outliers with $\omega_{\rm LO} > 300$ meV. All these systems have one or more hydrogen atoms,
leading to high frequency non-dispersive phonon modes (molecular-type vibrational levels). 
This alone does not necessarily result in large values of $\ga$, but in the presence of halides or chalcogens the valence band mass also tends to increase, which does lead to large values of ZPR. 
The second point is that materials with one or more halides are concentrated at the lower end of the distribution in phonon frequencies, but this is compensated by the other parameters which are electronic, leading to both large $\ga$ and ZPR. 
The conduction bands of these materials do not possess such heavy masses, and so the ZPR ends up being smaller.

A general trend can be derived from the datasets: the values of $\ga$ and ZPR will increase if more polar elements are present in the compound, especially halides. These produce large band effective masses and electronic dielectric constants, and so smaller $\epsilon^*$. 
Despite lower values of $\omega_{\rm LO}$, the $\ga$ in halides is often beyond the limit of validity of perturbation theory for the standard \fro model, indicating the probable breakdown of the 
first-principles AHC approach.

\begin{figure*}
\centering
\includegraphics[width=\linewidth]{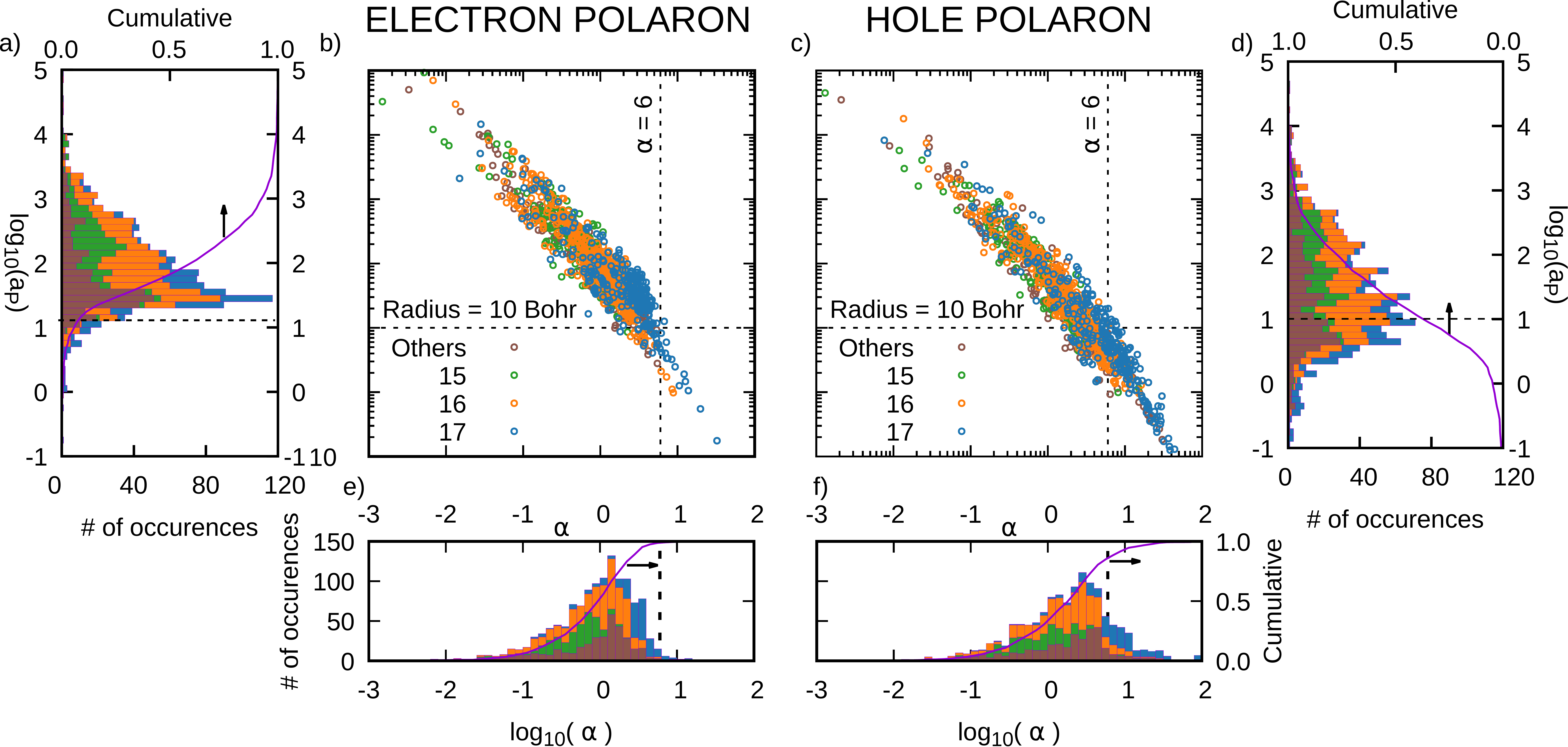}
\caption{Polaron radii and $\ga$ distributions for both holes and electrons within the standard \fro model. Dashed lines show the limits of \fro perturbation theory ($\ga=6$) and small polarons (indicatively $\mathrm{a_P}$ = 10 Bohr). Hole polarons are clearly heavier and more localized, but the majority of both distributions is within the limits of validity of the \fro model. Histograms show the statistical and cumulative distribution, with stacked bar graphs for the different chemical periods. See Table S.5 for the distribution of radii and coupling strength of both electron and hole polarons.
Same color code as in Fig.~\ref{fig:szpr-2}.
}
\label{fig:radii_alphas}
\end{figure*}

\begin{figure}
    \centering
    \includegraphics[width=0.9\columnwidth]{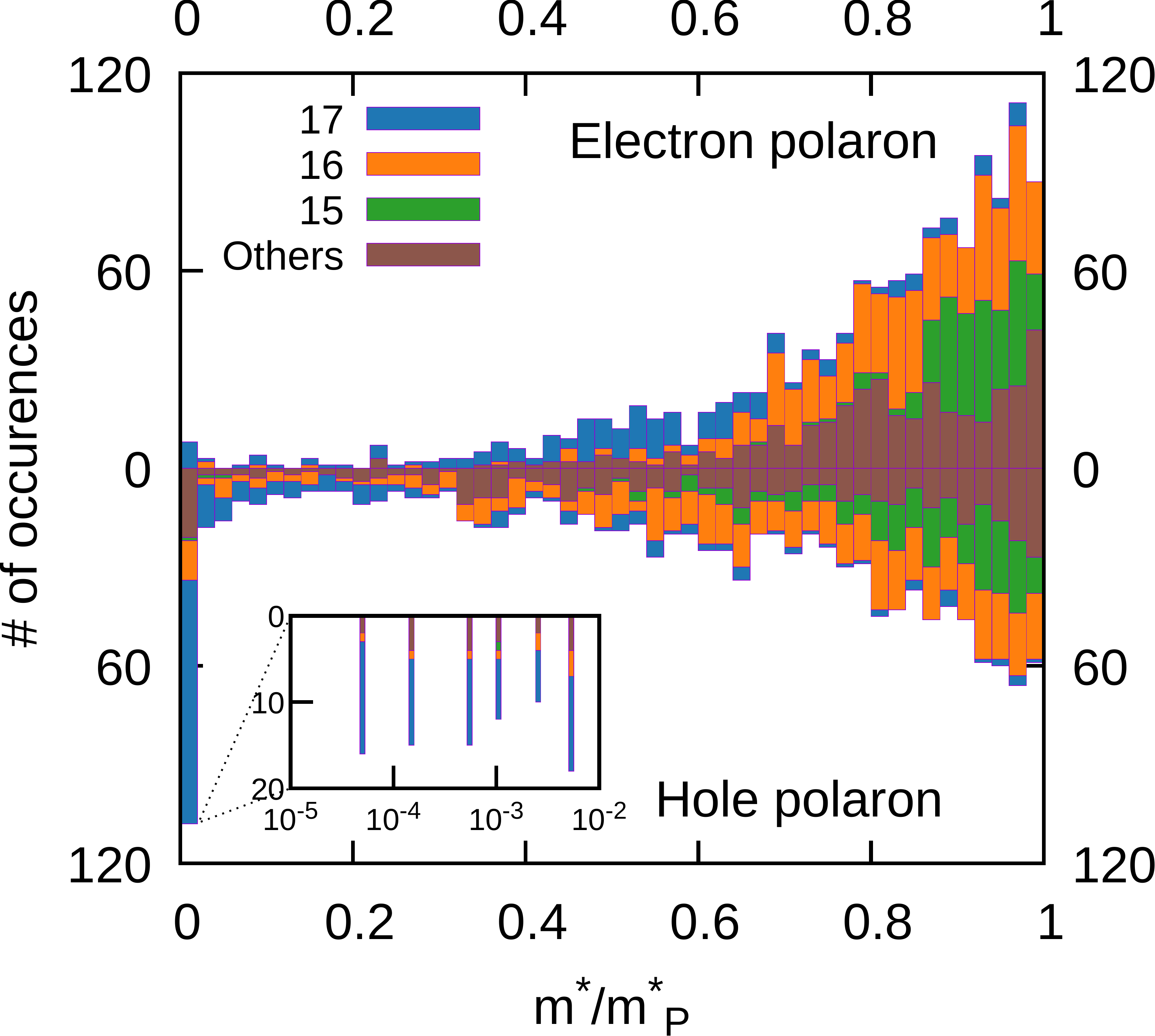}
    \caption{Inverse effective mass enhancement, for both hole (positive) and electron (negative) polarons within the standard \fro model. Stacked bars correspond to chemical period. The inset shows the distribution for very heavy hole polarons with huge mass enhancement. The effective mass enhancement is based on the mapping provided the Diagrammatic Monte Carlo results (see Fig. 5 in Ref. \emph{\citenum{Mishchenko2000}}) and Eq.~\eqref{eq:mp_dmc}. Same color code as in Fig.~\ref{fig:szpr-2}.  }
    \label{fig:eff_mass_enh}
\end{figure}

In addition to the analysis of the validity of perturbation theory thanks to limits on $\ga$, the validity of the large polaron hypothesis can also be assessed. 
This hypothesis is crucial for the \fro approach, 
be it in the standard form or in the generalized form.
For this purpose, the polaron radius, Eq.~\eqref{eq-ap-radius}, is computed, in the strong-coupling approximation.
Such information is combined with 
the $\ga$ data for both conduction and valence band edges in Fig.~\ref{fig:radii_alphas}. 
Histogram distributions of $\ga$ and $\mathrm{a_P}$ values
are shown. 
An indicative value of $\mathrm{a_P}=10$ Bohr has been
chosen to indicate the frontier between small polarons and large polarons. 
Similarly, and as already discussed, 
values of $\ga$ larger than 6 loosely indicate breakdown of 
perturbation theory.
Materials with small $\mathrm{a_P}$ will not be well reproduced with the long-range, large polaron \fro approximation.
See Table S.5 for the statistics of both electron and hole polarons. 
The number of cases yielding large polarons that can be described by perturbation theory is quite high: about 91\% of the materials for electron and 58\% for hole polarons.
The large polaron hypothesis breaks down for about 5\% of cases, for electrons and 34.5\% of the hole polarons.
The remainder, namely, materials for which the
large polaron hypothesis is valid, but for which perturbation
theory breaks down is very small : only 0.08\% for the
electron polarons, and 0.56\% for the hole polarons.

The EPI enhancement of the bare electronic effective mass is shown in Fig.~\ref{fig:eff_mass_enh}. The contribution is determined based on the diagrammatic quantum Monte Carlo calculation proposed by Mishchenko et al. (see Fig. 5 in Ref. \emph{\citenum{Mishchenko2000}}).
Considering the improved generalized \fro model, polaron anisotropy will shift many materials to lower critical radii, and the breakdown of perturbation theory can occur at lower $\ga$, as shown in Ref.~\onlinecite{Guster2021}, where a similar analysis of repartition of $\ga$ and $\mathrm{a_P}$
was performed for a much smaller set of materials, all
exhibiting cubic symmetry.

On this basis, the large polaron hypothesis with perturbative treatment might still be appropriate to treat the electron polaron for a majority of materials, but this might not be true to treat the valence band.
The breakdown of perturbation theory for $\ga$ bigger than 6, implying also
that first-principles AHC theory would be inappropriate,
is less often encountered, even for valence bands.

\subsection{Generalized \fro model}\label{sec:HT_generalizedmodel}

\begin{figure}[!hptb]

\subfloat{\includegraphics[width=0.9\columnwidth]{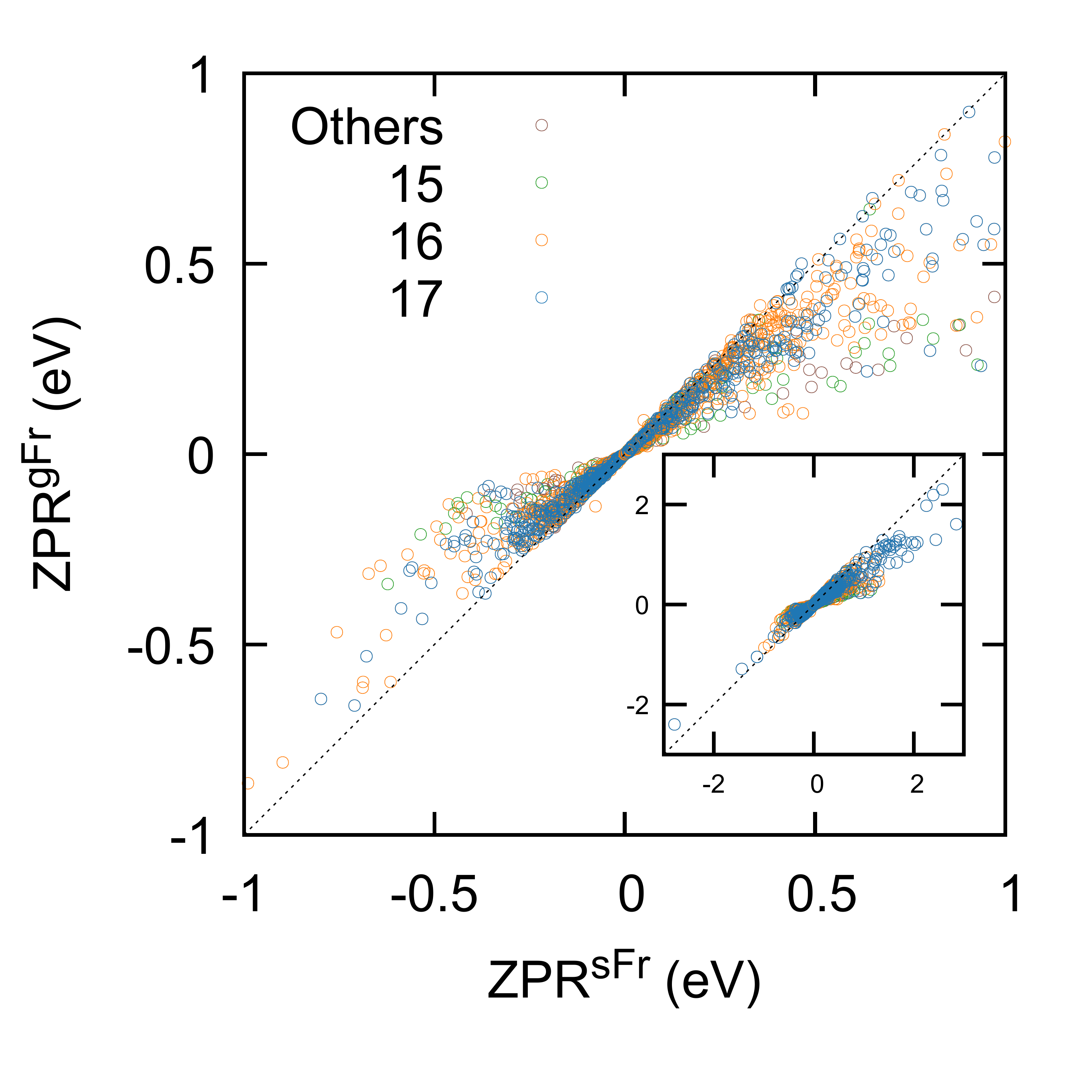}
\label{gzpr_vs_szpr_v}}
\hfil
\subfloat{\includegraphics[width=0.9\columnwidth]{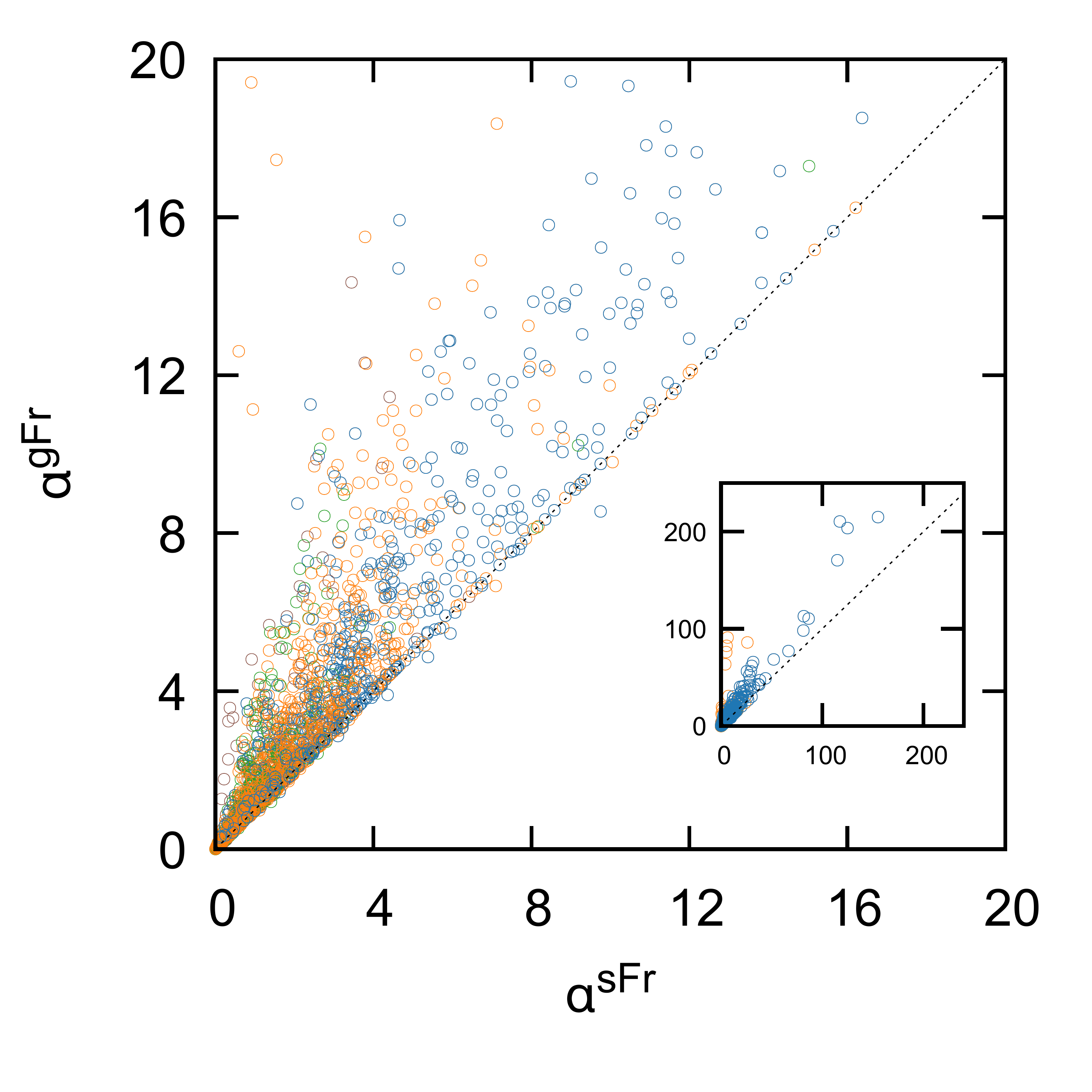}
\label{gzpr_vs_szpr_c}}
\caption{Comparison of generalized and standard \fro model for ZPR (top) and $\ga$ (bottom), for both valence and conduction edges.  
The insets show the dispersion for the full range of values of ZPR and $\ga$. 
The generalized ZPR is clearly smaller than the standard \fro model, often by a factor of 2 or more, whereas the $\ga_{\rm S}$ and $\ga_{\rm G}$ are quite close, with slightly larger $\ga_{\rm G}$ values. This shows that different factors influence the ZPR and $\ga$.}
\label{fig:ga_vs_sa}
\end{figure}

We now compare our results to the generalized \fro model discussed in Section~\ref{sec:gfro}. 
Here we define the direction-dependent dielectric tensor as $\epsilon^\infty(\hat \qq) = \sum_{\ga\beta} \hat q_\ga \epsilon^\infty_{\ga\beta}\hat q_\beta$,
we take into account all phonon modes with their respective
direction dependence $\omega_{j0}^*(\hat{\qq})$ and coupling, $\epsilon_j^*(\hat{\qq})$,
and the direction-dependent effective mass 
inside Eq.\eqref{eq-gzpr}.
To evaluate the directional dependence, all quantities that depend on $\hat \qq$ are computed on a sphere of radius $10^{-4}$ Bohr$^{-1}$, using a total of 2000 points to sample the sphere. 
The phonon frequencies and eigendisplacements are interpolated on this grid using the \emph{anadbb} tool in the ABINIT software package. Note that all parameters (e.q. k-point grids, energy cutoffs, pseudo potentials) are the same for the evaluation of all quantities involved. Differences that arise will come only from the nature of each method,  
namely that in the generalized model the angular dependency of all quantities is taken into account and the effects of all phonon modes are included, weighted by the mode-polarity vectors. 

For the generalized \fro model, no all-range calculations
of the Feynman or DMC type have been performed until now.
We have thus to rely on the perturbative result, Eq.~\eqref{eq-EP-gen}, to obtain the ZPR. 
Coherently, the lowest order of perturbation, Eq.~\eqref{eq-EP}, is used to compare with the standard \fro model.
For $\ga$ in the generalized \fro model, we employ the expression
\begin{equation}
    \ga_{\rm G} = \sum_j \braket{\ga_j(\hat \qq)}_{\hat \qq},
\end{equation}
with $\ga_j(\hat \qq)$ given by Eq.~\eqref{eq-ga1-jq}.

In Fig.~\ref{fig:ga_vs_sa} we show the comparison between the standard and the generalized \fro model ZPR, both obtained in low-order perturbation theory, that we denote ZPR$_{\rm S}$ and ZPR$_{\rm G}$, respectively, and $\ga$, that we denote $\ga_{\rm S}$ and $\ga_{\rm G}$, respectively.  While it is not apparent to the naked eye, ZPR values for binary cubic systems match in both the generalized and standard models (see the file binarycubicZPRcomparison.json provided as SI).
From Fig.~\ref{fig:ga_vs_sa} it is apparent that the generalized model reduces the value of the ZPR, in some cases drastically. 
For instance the valence ZPR goes from 3164.39 meV to 548.18 meV 
for Rb$_2$HBrO, a difference of 82\%, and the conduction ZPR of Li$_2$CaHfF$_8$ goes from 3484.27 meV to 2177.84 meV, a difference of 37\%. 
For Rb$_2$HBrO the reason for this drastic reduction comes from the mode polarity vectors, which re-weight the contribution of each phonon mode, and the vibrational molecular modes arising from the presence of hydrogen atoms see their contributions diminished, so the $\ga_j(\hat \qq)$ factor in Eq.~\eqref{eq-gzpr2} cancels the very high mode frequency. 
This also occurs for other outlying compounds of Fig.~\ref{fig:szpr-2}.

For the conduction bands, the relative reductions of the largest ZPR$_{\rm G}$ values are much smaller in comparison, with a 37\% reduction for Li$_2$CaHfF$_8$ and 39\% for K$_2$TiF$_6$, i.e. those with the largest \zprs{}. While some re-weighting happens thanks to the mode-polarity vectors, these materials have massive, almost point-defect-like, conduction bands. The main contribution to the reduction will then come from the fact that we account for the geometry of the Brillouin zone in the integration. 

Finally we look at how $\ga_{\rm G}$ compares to $\ga_{\rm S}$ in Fig.~\ref{fig:ga_vs_sa} b). The outlying material for both conduction and valence cases is CsNO$_2$, which has both heavy conduction and valence band masses and high-frequency phonon modes. 
However, the largest $\ga_j(\hat \qq)$ contributions that enter into Eqs.~\eqref{eq-gzpr2} and~\eqref{eq-ga1-jq} are from modes at much lower frequencies. 
While this delivers a large $\ga_{\rm G}$, it does not make this material an outlier in terms of \zprg, and illustrates how mode and direction independent quantities are not fully reliable and can fail in some cases.
Nevertheless, for all other materials the differences between $\ga_{\rm G}$ and $\ga_{\rm S}$ are within 50\% of $\ga_{\rm S}$.

Some interesting outliers were found when scanning through the values of ZPR and $\ga$ in both models for binary compounds, including a family of alkali metal nitrides with the chemical formula XN$_3$. These are discussed in the next sections, together with other materials which have large ZPR due to a high number of fluorine ions, comparing to benchmark results from a fully first-principles method. 

\section{Ab initio benchmarking}\label{sec:abichecks}

The generalized \fro model is expected to improve over the standard model.
In order to quantify the improvement brought by the generalized \fro model, 
we compare model results with fully ab initio (AHC) calculations of the ZPR.
As the latter are much more costly, we have selected a limited set of representative and/or simple test cases.
Like the generalized \fro model, the first-principles AHC approach works in the lowest order of perturbation.

\subsection{Results and comparison to Miglio et al.}

Previously, Miglio et al\cite{Miglio2020} computed ZPR from first principles for a set of 30 materials and compared them with the generalized \fro model.
Most of the stronger ionic compounds (oxides and chalcogenides) were well described by the model, within 25\% error compared to the first principles AHC approach.
For nitrides, the ZPR were less accurate but still within 50\% error. Their ZPR was twice larger in first principles calculations than using the generalized \fro model. 
We calculate the valence band ZPR for an intersecting subset of 20 materials (Tab. S.6) based on the generalized ZPR for cubic materials, Eq.~\eqref{eq-gzpr-cubic-simple}.

In addition, we have chosen specific systems with high ZPR, combined with either low, medium or high $\alpha$. 
Despite our theories being based on a perturbative
approach, the latter case is nevertheless instructive
as it is expected that the similarity or difference
within a common perturbative framework of similar order will translate to a similarity or difference within
more elaborate frameworks able to tackle non-perturbative
behaviors.
The first case of lower $\alpha$ are four ionic molecular crystal azides: KN$_3$ and RbN$_3$ crystallize in a tetragonal system, LiN$_3$ and NaN$_3$ in a monoclinic system; the following case contains the trigonal system CsNO$_2$ (medium $\ga$); and, for the final extreme case we examine cubic Cs$_2$NaScF$_6$, tetragonal Li$_2$CaHfF$_8$ and trigonal K$_2$TiF$_6$.

\begin{figure}[!htpb]
    \centering
    \includegraphics[width=\columnwidth]{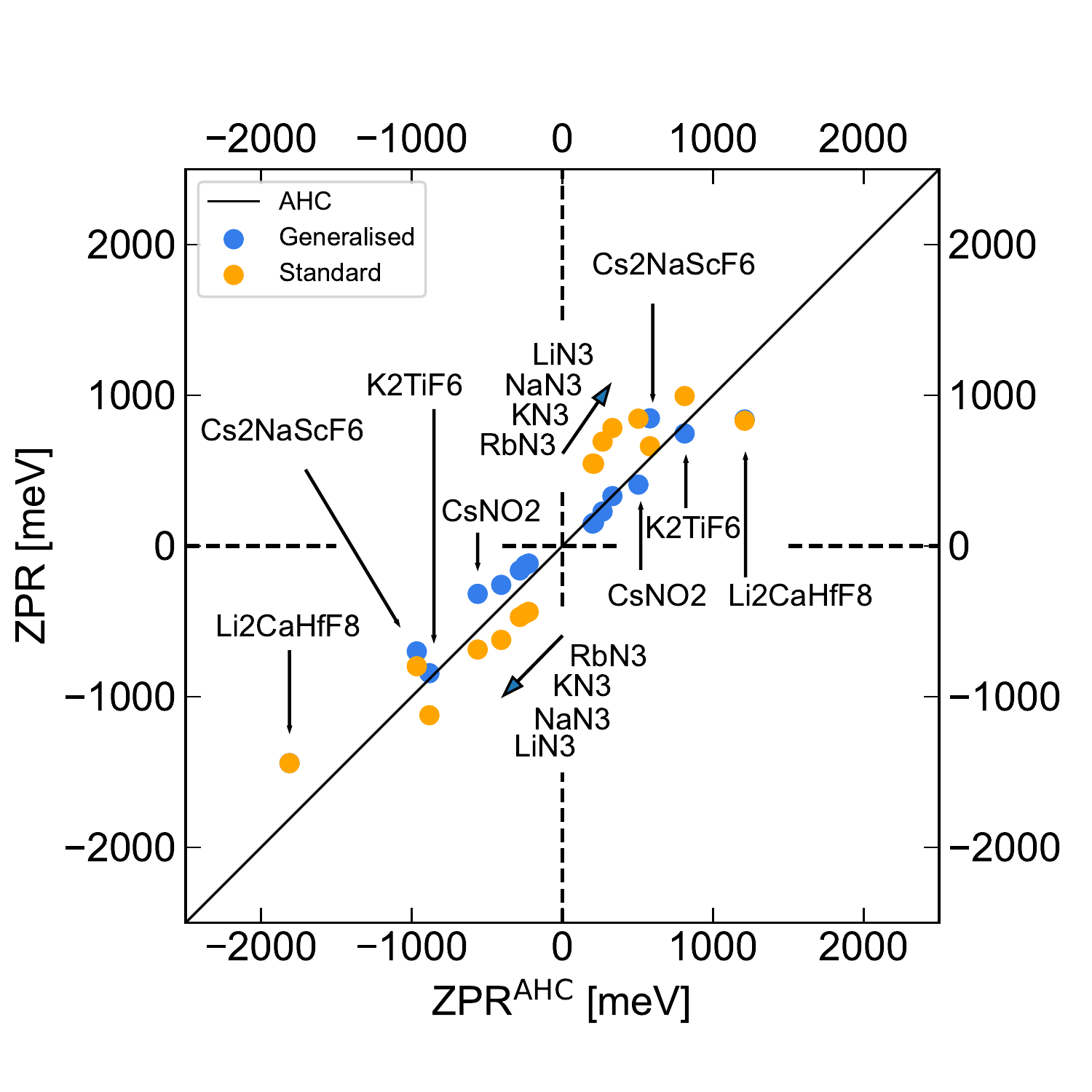}
    \caption{Comparison of the VBM (positive, right top part) and CBM (negative, left bottom part) ZPRs between the standard and generalized \fro models and the non-adiabatic AHC approach for some chosen systems. 
    }
    \label{fig:ZPR_Azides_Cs}
\end{figure}

In Fig.~\ref{fig:ZPR_Azides_Cs} we see that overall the standard \fro model overestimates the non-adiabatic AHC ZPR, while the generalized \fro model underestimates it. 
The former relies solely on the contribution from the highest $\omega_{LO}$ phonon branch. 
The latter accumulates the contribution to the ZPR from all the LO-phonon modes, and the properties are averaged for all directions $\mb{q} \to 0$. 

There are still important contributions missing from the generalized \fro model, such as the non-polar and TO phonon modes, but the trend is surprisingly strong in both cases: the dominant qualitative physics is already present in the standard \fro model, and, in most of the cases, the generalized version quantitatively corrects the ZPR. 

The systems Cs$_2$NaScF$_6$ and Li$_2$CaHfF$_8$ are the only cases in which the calculated values do not all follow this global trend. Their values in the standard and generalized models are very close due to a low $\omega_{\rm LO}$. Even if the ZPR is spread over other phonon modes, their frequency would not be far from the low $\olo$. In SI Table S. 2, we average $\omega$ over the phonon modes using the ZPR as weight. The farther this average is from $\olo$, the higher is the importance of phonons at lower frequencies

The main parameters of the standard and generalized \fro approaches are shown in SI Table S. 1 and in Table S. 2, respectively,
for our benchmark materials. 

In the following Secs. \ref{subsec:azides} and \ref{subsec:Cscompound} for KN$_3$ and Cs$_2$NaScF$_6$,  we determine the features which produce differences between the standard and generalized \fro approaches, and with the non-adiabatic AHC method.

\subsection{Azides: large ZPR, small $\alpha$}\label{subsec:azides}

For the azide series, the highest $\omega_{\rm LO}$ are similar with a maximum difference of 14 meV.
Therefore, the changes in the ZPR come from the dielectric constant and the effective mass. The trend from Rb, K, Na to Li is a decrease of ionic radius and an increase of electronegativity\citep{Gordy1946}, leading to a decrease of the unit cell volume and the dielectric constant, and an increase of the effective mass. 

In the standard \fro model the value of $\alpha$ of KN$_3$ is smaller than, for example, the KF system. Even if both have a similar fraction $\sqrt{m^*} / \epsilon^*$ (0.25 for KF and 0.28 for KN$_3$) KN$_3$ has a higher $\omega_{\text{LO}}$ (271 meV), see SI Table S.3 or SI Fig. S.4) than KF (42 meV). The very high $\omega_{\text{LO}}$ of KN$_3$ comes from the resonances created by the linear chains of N$_3$ harmonic oscillators, where all three atoms vibrate along the bonds. The same reasoning can be applied to the other azides.

In our choice of parameters for the standard \fro model, the system LO phonon mode with the highest frequency is selected as the sole contributor to ZPR, which is not the case for the generalized \fro model. 
In the case of KN$_3$, there are two active LO phonon modes in the  generalized \fro approach (see bottom of SI Table S.3).
The phonon mode with highest contribution (76\%) to the ZPR is LO phonon mode j=15 with a frequency of 20 meV, and not the highest LO mode (j=24) with a frequency of 260 meV.
As the contribution for the ZPR changes from the highest LO mode to a lower one, the value of $\alpha$ also changes as it is explicitly proportional to the phonon frequency by $\omega_{\text{LO}}^{-1/2}$ both for standard and generalized Fr\"ohlich, Eqs. \eqref{eq-alpha} and \eqref{eq-ga1-jq} respectively.
The renormalization including all LO modes translates a low $\alpha = 2.03$ for the standard \fro model into a higher $\langle \, \alpha_j (\hat{\qq}) \rangle_{j\hat{\qq}} = 5.09$ using the generalized \fro model. The value of $\alpha$ can nonetheless also be lower in a few cases for the generalized \fro model (see Fig.~\ref{fig:ga_vs_sa}), as the re-distribution of the ZPR to other phonon modes can be compensated by effective mass anisotropy or band degeneracy effects.

Going beyond the generalized \fro approach and deconstructing the non-adiabatic AHC ZPR into its phonon mode components (see top of SI Table S.3), we find that the highest contribution of 26.14\% to the total ZPR comes from LO phonon mode 15 and that there are non-LO phonon modes (20, 21 and 22, excluded from the \fro models) which have even higher contribution to the ZPR than the highest frequency LO phonon mode 24. 

Several factors contribute to the spread of the ZPR contribution throughout the different phonon states.
One key attribute is the polarizability and eigenvectors of the modes. Phonon mode 15 in KN$_3$ has a stronger polarizability than the highest phonon mode, which is mostly driven by the lightest atoms. Mode 15 has a mix of contributions between the potassium and nitrogen atomic vibrations, with larger dipoles, and hence a larger mode polarizability.

\begin{figure}[!hptb]
    \centering
    \includegraphics[width=\columnwidth]{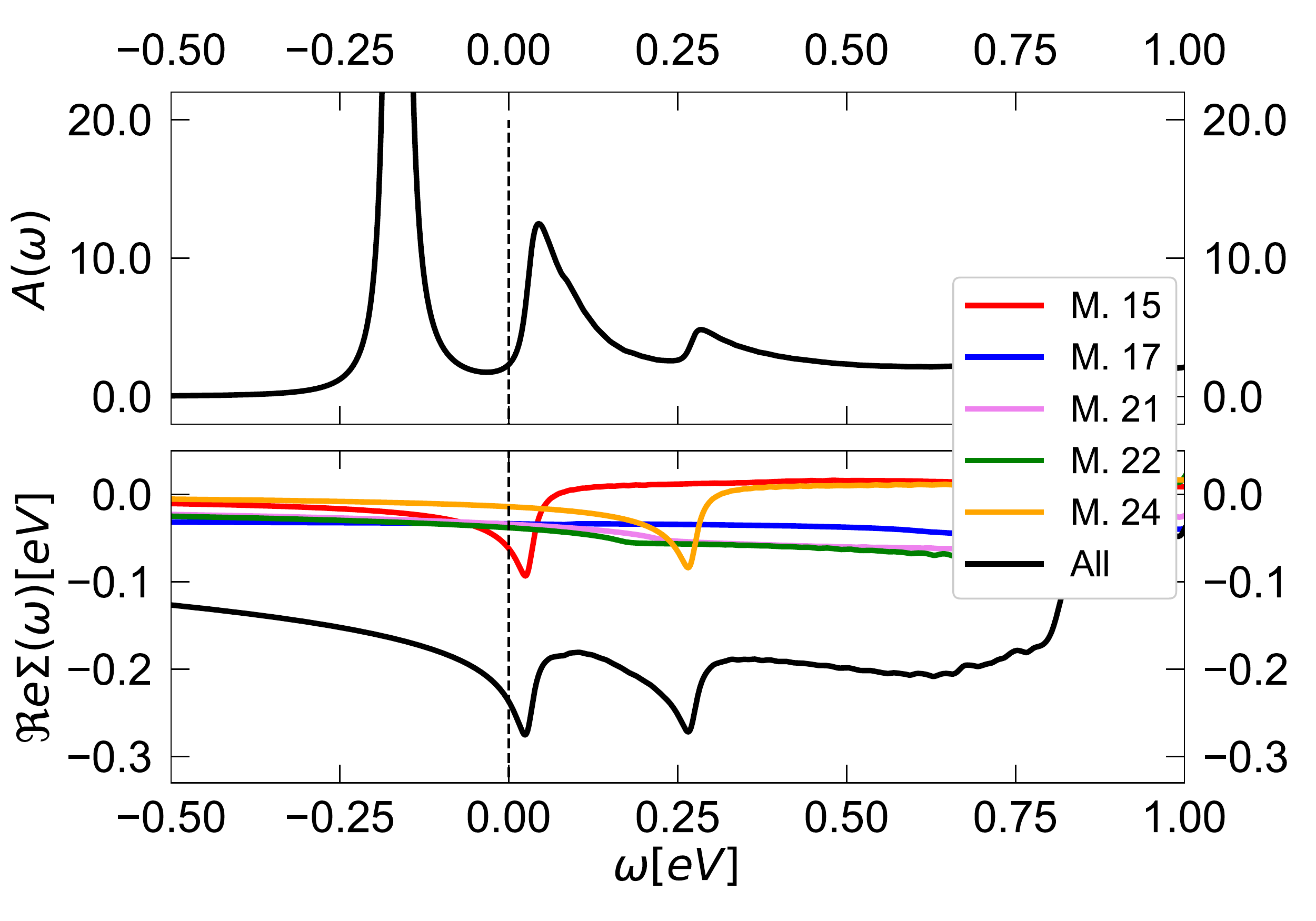}
    \caption{Spectral function, A($\omega$) (top) and the real part of the self-energy, $\Re{e}\Sigma(\omega)$ (bottom) for KN$_3$ at the CBM. The self-energy is split by phonon mode for the largest contributions to the ZPR$_c$ as in SI Table S.3.}
    \label{fig:KN3_sigmaaw}
\end{figure}

A second key factor is the inclusion of non-LO modes. In Fig. \ref{fig:KN3_sigmaaw}, we show the electron self-energy of KN$_3$ (bottom) and the spectral function (top). The LO-phonons can be found in the self-energy with characteristic \fro peaks\cite{Nery2018}. Specifically in KN$_3$, the self-energy terms for phonon modes 15 and 24 show a peak at their $\Gamma$ point phonon frequency (the KS energy is at 0.0 eV). Mode 17 is also LO, but the peak amplitude is small, around 1 meV.  In addition to their contribution to the ZPR binding energy, the LO-phonon modes are also responsible for satellites in the spectral function visible in the top panel. 

The non-LO phonon modes (either TO or non-polar) do not show peaks but a type of plateau starting at their $\Gamma$ point phonon frequencies, linked to long-range quadrupole potential\cite{Brunin2020,Brunin2020a} and/or short-range fields\cite{Abreu2022}. At the KS energy, where the ZPR is evaluated, their $\Re e \Sigma(\varepsilon)$ is not negligible at all, with a net contribution larger than that of the LO modes.

\begin{figure}[h!]
    \centering
    \includegraphics[width=\columnwidth]{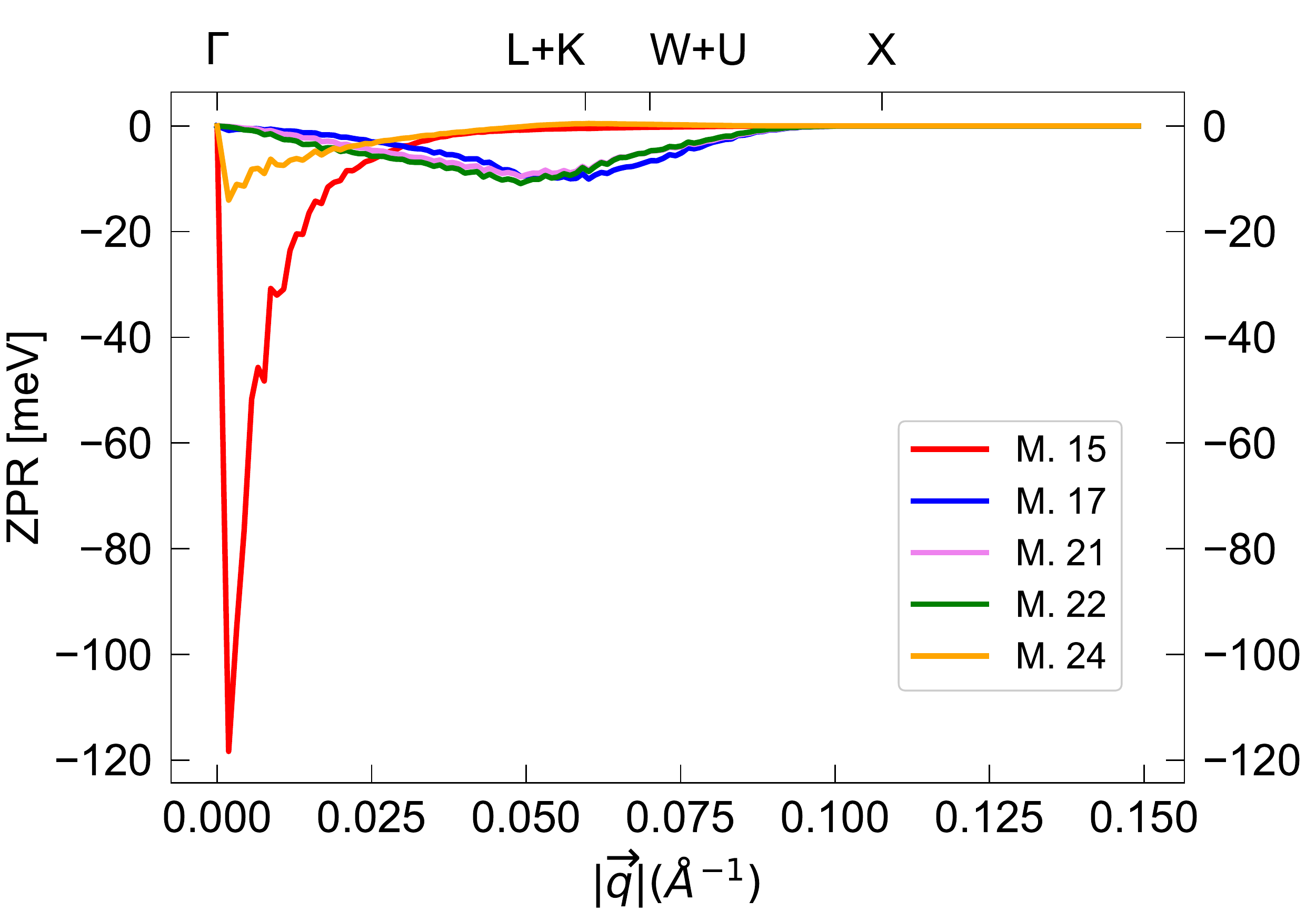}
        \caption{Spherical accumulation of the ZPR$_{\rm c}$ of KN$_3$ as a function of the norm of the vector $|\mb{q}|$ for each phonon mode using the $64 \times 64 \times 64$ $\mb{q}$-grid sampling. The norms of selected high symmetry points are shown on the top axis.}
    \label{fig:KN3_zprnormq}
\end{figure}

Another way to distinguish contributions to the ZPR is by plotting their dependency on the phonon wave-vector (norm), as in Fig. \ref{fig:KN3_zprnormq}. The two LO phonon modes have their main contributions from wave-vectors close to $\Gamma$, and correspond to long-range electric dipole fields. The non-LO phonon modes (j=17, 21, and 22) originate at the boundary of the Brillouin zone, and correspond to interactions with shorter-range crystal fields. We note that it is important in this analysis to avoid mixing LO and non-LO band contributions when their frequencies cross away from $\Gamma$, by following the irreducible representations and character of each mode to attribute the ZPR$(j, \qq)$ contributions.

\subsection{Cs$_2$NaScF$_6$: large ZPR, large alpha}\label{subsec:Cscompound}

The ZPR$_\text{c}$ of the \Cs compound is much higher than the azides with a value of -966.1 meV in the non-adiabatic AHC. In addition to the small increase of the effective dielectric constant contribution, there is also an approximately 8-fold increase of the effective mass compared with the azides (SI Table S. 1). The \Cs electron band structure and the projected density of states are shown in SI Fig. S.5. The bottom conduction band has very low dispersion which translates into a large effective mass and localized electrons, which are found in the Sc-F bonds. The main source of the high ZPR is the coupling between the (high frequency) vibrations of F atoms and the d-orbital conduction band of Sc.

Dissimilar to the azides, both the standard and the generalized \fro models underestimate the non-adiabatic AHC ZPR$_\text{c}$ of the \Cs compound by 168 and 267 meV, respectively. Surprisingly, the generalized \fro model gives worse result than the standard when comparing to the non-adiabatic AHC approach, through a compensation of errors between neglecting lower frequency LO modes and neglecting non-LO modes altogether.
The ZPR calculated within the generalized \fro and non-adiabatic AHC can be split into phonon mode contributions (SI Table S.4). 
Unlike the case of the azides, the highest ZPR contribution is the non-LO phonon mode 29, contributing almost 45\% and an $\alpha_i$ of 9.20. Generalized \fro ignores the non-LO phonons, which, in this case, are close to the $\olo$. Part of this contribution is spread to lower LO phonon modes leading to a worse ZPR.

\begin{figure}[!hptb]
    \centering
    \includegraphics[width=\columnwidth]{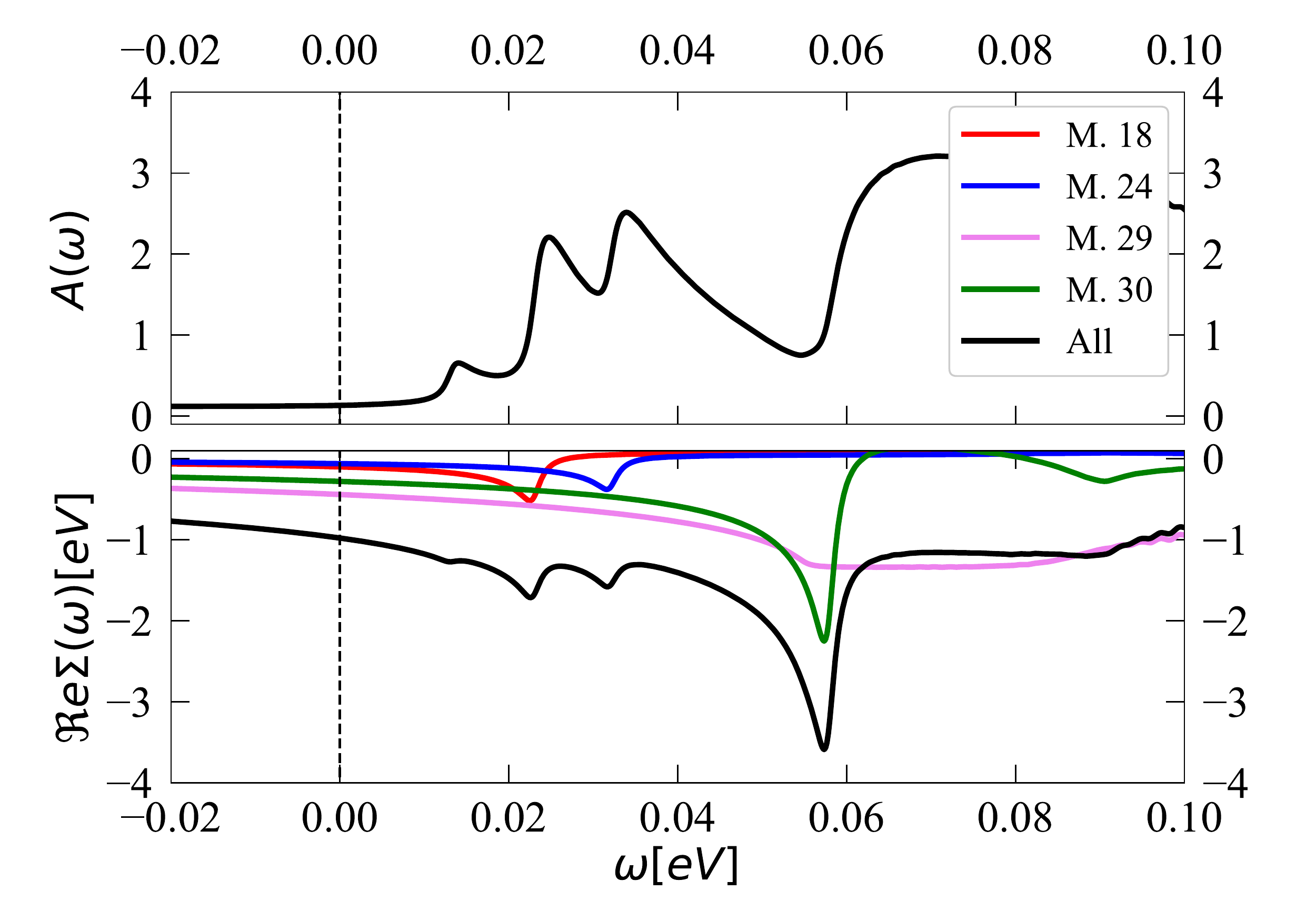}
    \caption{Spectral function, A($\omega$) (top) and the real part of the self-energy, $\Re{e}\Sigma(\omega)$ (bottom) for Cs$_2$NaScF$_6$ at the CBM. The self-energy is split by phonon mode for the largest contributions to the ZPR$_{\rm c}$ as in SI Table S.4. The frequency range is limited to the satellites: the two QP peaks at -0.249 and 0.341 eV are not visible in this range. The first peak in A($\omega$) comes from a mode with low contribution to the full ZPR, which is only a small bump in the total $\Re e\Sigma(\omega)$.
    }
    \label{fig:Cs2NaScF6_sigmaandaw}
\end{figure}

The real part of the self-energy for phonon mode 29 is the most important at the KS energy (0.0 in Fig. \ref{fig:Cs2NaScF6_sigmaandaw}), and increases towards its $\Gamma$-point phonon frequency (47.8 meV) but it has no peak. The phonon mode with the highest frequency is a LO phonon mode, showing a green peak in the self-energy figure, but has a smaller contribution of around 28.5\% to the total ZPR. The other phonon modes shown in the figure are also LO phonon modes and have small self-energy peaks at their $\Gamma$ point phonon frequencies. The spectral function shape is complicated by the presence of two quasi particle solutions (not visible in the figure), which convolute the full self-energy. This is more common in the electron-phonon case (as opposed to electron-electron) as the self-energy amplitude is of the same order of magnitude as the phonon energies (or even larger).

\begin{figure}[!hptb]
    \centering
    \includegraphics[width=\columnwidth]{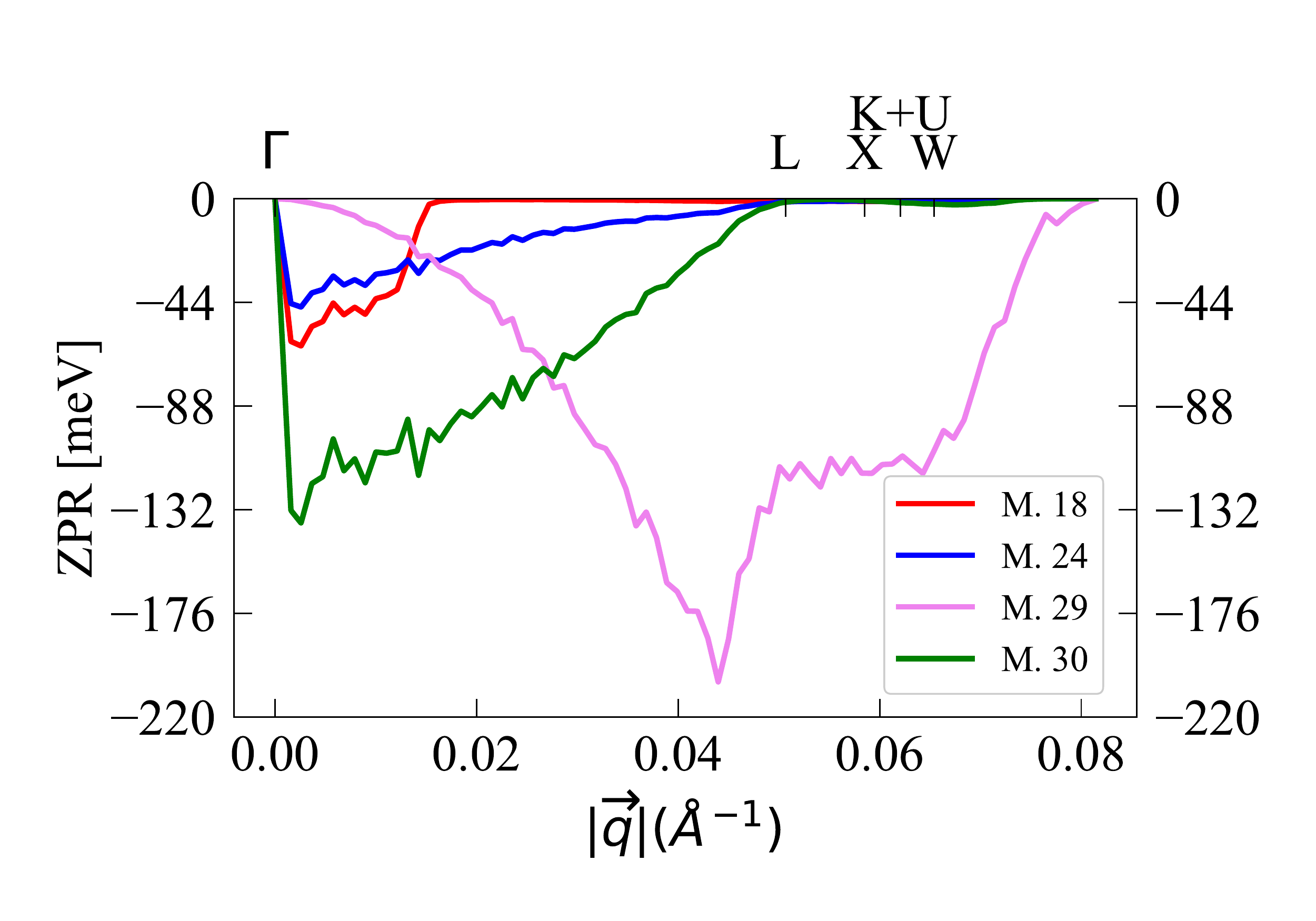}
    \caption{Spherical accumulation of the ZPR$_c$ of Cs$_2$NaScF$_6$ as a function of the norm of the vector $|\mb{q}|$  for each phonon mode using the $64 \times 64 \times 64$ $\mb{q}$-grid sampling.}
    \label{fig:Cs2NaScF6_ZPRnormq}
\end{figure}

The angle-integrated ZPR as a function of wave-vector norm (Fig.~\ref{fig:Cs2NaScF6_ZPRnormq}) shows the Fr\"ohlich-like behavior close to $|\mb{q}|=0$ for $j=18, 24$, and $30$. 
Mode $29$ behaves as $|\mb{q}|^2$ following the volume contribution $4 \pi |\mb{q}|^2$ in the angular average, which means the ZPR contributions are relatively constant throughout the Brillouin zone.

As a summary for the set of ``extreme'' materials considered this section, we observe that the standard and generalized \fro models are both close to the full first principles trend, but overestimate and underestimate, respectively, the non-adiabatic AHC ZPR. One characteristic of high $\olo$ materials, as in the azides, is the geometrical isolation of the lighter atoms. If lower frequency phonon modes have high polarization, the generalized \fro ZPR is spread over these modes, reducing the ZPR compared with the standard \fro model, which includes solely the highest $\olo$ phonon mode. This spread also leads to an increase of the averaged $\alpha$ value, as the electrons interact with lower frequency phonon modes. In cases with intrinsically low $\olo$ phonon modes, the ZPR can be similar for both models. In addition, ignoring non-LO modes can worsen the generalized \fro model results, especially if the non-LO phonon mode is close to $\olo$, exaggerating the importance of lower frequencies and widening the distance from the non-adiabatic AHC ZPR.

\section{Conclusions}\label{sec:conclusions}

We evaluate the polaron binding energy, or zero point renormalization, in both standard and generalized \fro models for a database of \numcompounds materials. 
Lowest order perturbation theory is used for the generalized \fro model, while both perturbative and all-range formulas are available for the standard model, once the single parameter $\ga$ is defined. 

In our study we find a broad range of validity for both models (58\% of valence bands and 91\% of conduction band polarons), but the generalized model is in better quantitative agreement with our fully ab initio spot checks using the Allen-Heine-Cardona theory. 
The generalized \fro model ZPR shows a significant decrease of the ZPR but a slight increase of $\ga$ with respect to the standard model. 
We find distinctive trends depending on the material's composition: more electronegative compounds containing halides or chalcogens generally present higher ZPR associated with higher $\alpha$, while compounds containing group 13 elements show rather lower ZPR, though there are outliers in all categories.

Given the broad range of behaviors in our set of materials, we complement our studies with fully ab initio DFT-based non-adiabatic AHC calculations of the ZPR. 
We focus on few outlier materials with different $\alpha$ (low, medium, and high) and high ZPR. Both \fro models follow the trend of the non-adiabatic AHC qualitatively. The lower and medium $\alpha$ materials show a quantitative improvement going from standard to generalized \fro, thanks in particular to the redistribution of the ZPR to lower frequency phonon modes. In some exceptional cases (e.g. Cs$_2$NaScF$_6$)  the standard \fro can give better (i.e. closer to AHC) results than the generalized model, when important non-LO phonon modes are ignored.

The standard \fro model can fail in more than one way: due to essential non-LO phonons modes, anisotropy, or, crucially, the breakdown of perturbation theory. 
However, regardless of the \fro method's limitations, we provide strong evidence for the ubiquity of polaron formation, the range of possible behaviors and parameter space, and the importance of polarons in providing reliable band gaps and effective masses. In a very small number of weak coupling cases the estimated polaron radius is small enough to call into question the applicability of \fro type models.

\begin{acknowledgments}
This work has been supported by the Fonds de la Recherche Scientifique (FRS-FNRS Belgium) through
the PdR Grant No. T.0103.19 - ALPS.
ZZ and PMMCM acknowledge financial support by the Netherlands Sector Plan program 2019-2023.
This project has received funding from the European Union’s Horizon 2020 research and innovation program under grant agreement No. 951786 - NOMAD CoE.
Computational resources have been provided by the CISM/UCLouvain and
the CECI funded by the FRS-FNRS Belgium under Grant No. 2.5020.11,
as well as the Tier-1 supercomputer of the F\'ed\'eration Wallonie-Bruxelles, funded by the Walloon Region under grant agreement No. 1117545. 
We acknowledge a PRACE award granting access to MareNostrum4 at Barcelona Supercomputing Center (BSC), Spain (OptoSpin project id. 2020225411).
Moreover, we also acknowledge a PRACE Tier-1 award in the DECI-16 call for project REM-EPI on Archer and Archer2 EPCC in Edinburgh.
\end{acknowledgments}

\bibliographystyle{unsrt}
\bibliography{literature}

\end{document}